\documentclass[12pt,oneside,onecolumn,floatfix]{article}
\usepackage{epsfig}
\usepackage{amsmath,amssymb,amsfonts}
\usepackage{a4wide}
\usepackage[top=30pt,bottom=30pt,left=48pt,right=46pt]{geometry}
\usepackage{slashed, enumitem}
\usepackage[utf8]{inputenc}
\usepackage{jcappub}
\usepackage{enumerate}
\usepackage{float,color}
\usepackage{subfigure}
\usepackage{multirow}
\usepackage{slashed}
\usepackage{placeins}
\usepackage{wasysym} 
\usepackage{mathrsfs} 
\usepackage{caption}
\usepackage{amsmath}
\usepackage{amssymb}
\usepackage{graphicx}
\usepackage{slashed}
\usepackage{multirow}
\usepackage{placeins}
\usepackage[dvipsnames]{xcolor}
\usepackage{epstopdf}
\usepackage{soul}
\usepackage{tikz}
\usepackage[capitalise, english]{cleveref}
\usepackage{siunitx}
\usepackage{xspace}
\usepackage{booktabs}
\usetikzlibrary{trees}
\usetikzlibrary{decorations.pathmorphing}
\usetikzlibrary{decorations.markings}
\usepackage[colorlinks=true,citecolor=blue,linkcolor=blue]{hyperref}

\definecolor{americanrose}{rgb}{1.0, 0.01, 0.24}
\definecolor{ao}{rgb}{0.0, 0.0, 1.0}

\newcommand\myshade{80}
\colorlet{mylinkcolor}{violet}
\colorlet{mycitecolor}{americanrose}
\colorlet{myurlcolor}{ao}

\hypersetup{
	linkcolor  = mylinkcolor!\myshade!black,
	citecolor  = mycitecolor!\myshade!black,
	urlcolor   = myurlcolor!\myshade!black,
	colorlinks = true
}

\usepackage{tikz,xcolor,hyperref}

\definecolor{lime}{HTML}{A6CE39}
\DeclareRobustCommand{\orcidicon}{\hspace{-3mm}
	\begin{tikzpicture}
	\draw[lime, fill=lime] (0,0) 
	circle [radius=0.16] 
	node[white] {\hspace{0.2mm}{\fontfamily{qag}\selectfont \tiny ID}};
	\draw[white, fill=white] (-0.0675,0.1) 
	circle [radius=0.01];
	\end{tikzpicture}
	\hspace{-5mm}
}

\foreach \x in {A, ..., Z}{\expandafter\xdef\csname orcid\x\endcsname{\noexpand\href{https://orcid.org/\csname orcidauthor\x\endcsname}
		{\noexpand\orcidicon}}
}

\graphicspath{{figs/}}

\title{Fast Flavor Oscillations of Astrophysical Neutrinos with 1,\,2,\,\ldots,\,$\infty$ Crossings}

\author{Soumya Bhattacharyya\orcidA{}}
\emailAdd{soumyaquanta@gmail.com}

\author{and Basudeb Dasgupta\orcidB{}}
\emailAdd{bdasgupta@theory.tifr.res.in}
\affiliation{Tata Institute of Fundamental Research,\\ Homi Bhabha
	Road, Mumbai 400005, India}

\abstract
{In the early Universe, as well as in supernovae and merging neutron stars, neutrinos have such high densities that they affect each other and exhibit collective flavor oscillations. A crucial ingredient for fast collective flavor oscillations is that the electron lepton number (ELN) distribution changes its sign as a function of direction, i.e., has a zero crossing.  We present a study in two dimensions and show how fast flavor oscillations depend on the ELN and its crossings. We show that a large number of crossings can inhibit flavor oscillations. This may be a natural self-limiting mechanism that stabilizes the flavor content of the dense neutrino gas in a vast majority of scenarios, especially the early Universe, where the angular distributions for all flavors are very similar and crossings occur mainly due to fluctuations.}

\subheader{TIFR/TH/20-47}


\begin{document}
\maketitle

\section{Introduction}
\label{sec:intro}
Astrophysical neutrinos are a valuable probe of fundamental physics and astrophysics. The focus of this paper is the so-called fast collective flavor evolution of neutrinos, which can have a dramatic effect in dense astrophysical environments, e.g., on the explosion of core-collapse supernovae, nucleosynthesis in supernovae and neutron star mergers, and perhaps for neutrinos in cosmology.

Neutrino oscillation in ordinary matter is governed through two frequency scales -- one is the vacuum oscillation frequency, $\omega_{E} =|\Delta m^{2}|/(2E)$, related to the mass-squared difference between two different mass eigenstates, $\Delta m^2 = m_2^2 - m_1^2$, and the other is the matter potential, $\lambda=\sqrt{2}G_{F}n_{e}$, coming from the coherent forward scatterings of neutrinos with electrons with possibly spatially varying density $n_e$ in the medium.\footnote{The quantum mechanical amplitude of forward scattering interferes with free propagation and gives a potential for flavor oscillations. Momentum and number changing collisions do not produce interference effects.} Broadly, high density, $\lambda\gg\omega_E$, suppresses flavor mixing. As a result, large flavor conversion in ordinary matter can occur either after $\lambda$ drops below $\omega_{E}$, whence ordinary neutrino oscillations ensue, or when these two frequency scales match, i.e., $\lambda \approx \omega_{E}$, giving rise to the well known phenomenon  of matter-enhanced oscillations~\cite{PhysRevD.17.2369, Mikheev:1987qk}. Thus one might think that neutrinos cannot oscillate deep inside stars or in the early Universe.
 
Neutrino-neutrino scattering in a neutrino gas gives rise to a new scale, $\mu=\sqrt{2}G_{F}n_{\nu}$, proportional to the neutrino density $n_\nu$~\cite{Pantaleone:1992eq}. This new potential makes the flavor evolution nonlinear, allowing novel collective flavor oscillations in dense environments. A salient feature of these collective oscillations is that neutrinos of different energies oscillate approximately at the same rate, i.e., the average $\omega_E$ for synchronized oscillations~\cite{Kostelecky:1994dt}, $\sqrt{\omega_E\mu}$ for bipolar or slow collective oscillations~\cite{Pastor:2001iu,Duan:2006an}, and as  rapidly as $\mu$ for fast collective oscillations~\cite{Sawyer:2005jk}. The slow variant of these collective oscillations was studied deeply starting from the mid-2000s, exposing a sequence of new effects such as swapping of the flavor-dependent spectra as well as momentum, space, and time-dependent flavor transformations~\cite{Hannestad:2006nj,Raffelt:2007cb,Fogli:2007bk,EstebanPretel:2007ec,Dasgupta:2007ws,Dasgupta:2009mg,Duan:2008za, Dasgupta:2008cd,Raffelt:2013rqa, Mangano:2014zda,Dasgupta:2015iia,Abbar:2015fwa}.

Fast flavor oscillations became a topic of wide interest only around 2015, starting with the influential paper by Sawyer~\cite{Sawyer:2015dsa}, which pointed out that the angular distributions for the different neutrino flavors are different in the decoupling region in a supernova and that can cause oscillations with rate $\propto\mu$. This was followed by two crucial studies which developed theoretical understanding of fast oscillations in the linearized regime~\cite{Chakraborty:2016lct,Dasgupta:2016dbv}. The first elucidated an underlying instability, showing how it has little dependence on $\omega_E$~\cite{Chakraborty:2016lct}, and the second clarified the role of homogeneity and stationarity, and showed that fast oscillations may occur for the the presumably realistic neutrino distributions inspired by state-of-the-art simulations~\cite{Dasgupta:2016dbv}. Subsequently, this subject has seen significant progress~\cite{Dasgupta:2017oko,Abbar:2017pkh,Dasgupta:2018ulw,Capozzi:2018clo,Abbar:2018beu,Capozzi:2019lso,Martin:2019gxb,Johns:2019izj,Capozzi:2020kge,Johns:2020qsk}, and it appears that fast oscillations can occur in  many environments~\cite{Wu:2017qpc,Morinaga:2019wsv,DelfanAzari:2019tez,Abbar:2019zoq,Glas:2019ijo,George:2020veu,Abbar:2020qpi,Hansen:2020vgm} and drastically alter the flavor composition therein~\cite{Bhattacharyya:2020dhu,Bhattacharyya:2020jpj}. In particular, they may lead to almost complete flavor depolarization~\cite{Bhattacharyya:2020dhu,Bhattacharyya:2020jpj}, limited by conservation of lepton asymmetry.

Properly understanding these new effects requires solving a set of coupled nonlinear partial integro-differential equations in a $3$ (spatial) + $3$ (momentum) + $1$ (time) = 7 dimensional space, which has remained unachievable till date.  As a result, most studies assume a high degree of symmetry and limit the flavor evolution to be along a single coordinate --- be it temporal evolution of a homogeneous gas or the spatial evolution of a stationary dense neutrino gas. Previous experience with slow collective oscillations~\cite{Raffelt:2013rqa, Mangano:2014zda,Dasgupta:2015iia,Abbar:2015fwa} suggests that such restrictions artificially exclude allowed modes of flavor evolution. Calculations in 1+1+1 dimensions are  superior~\cite{Bhattacharyya:2020dhu,Bhattacharyya:2020jpj}, but as we will discuss in Sec.\,\ref{sec:truevsapp} they too are not fully satisfactory. While a calculation in 3+3+1 dimensions remains the holy grail, going to at least 2+2+1 dimensions is necessary to satisfy the constraints on neutrino velocities, include fully self-consistent space-time evolution, and not artificially exclude allowed modes of flavor evolution.

In this paper, we present a study of fast flavor oscillations of a dense neutrino gas in 2+2+1 dimensions. Our study explores how the flavor evolution is related to the symmetries and the number of zero crossings of the electron lepton number (ELN) distribution, i.e., the difference of $\nu_e$ and $\bar\nu_e$ angular distributions; a more precise definition appears later. We perform detailed comparisons of the numerical results with those from a linearized analysis, finding excellent agreement. The main insights obtained from our study are that the flavor evolution inherits symmetries of the ELN, and why ELNs with a large number of zero crossings can in fact lead to a much weaker flavor instability. 

The paper is organized as follows: Sec.\,\ref{sec:eoms} sets up the problem. Sec.\,\ref{sec:elntype} introduces the types of neutrino angular distributions that we consider in this study. In Secs.\,\ref{sec:lsa}, \ref{sec:numstrat}, and \ref{sec:anastrat}, we describe the analytical and numerical methods we use for solving the equations. In Sec.\,\ref{sec:res}, we discuss our numerical and analytical results for all examples in a systematic way. In Sec.\,\ref{sec:conc}, we conclude with a brief summary.

\section{Set-up and Methods}
\label{sec:setup}

\subsection{Equations of Motion}
\label{sec:eoms}
Consider an effective two-flavor framework with each neutrino being a superposition of the $e$ and $\mu$ flavors. Neglecting momentum-changing collisions, the flavor evolution of the neutrinos at position $\vec{x}$ and time $t$, with momentum $\vec{p}\approx E\vec{v}$, is given by~\cite{Sigl:1992fn}
\begin{equation}\label{1}
\begin{split}
\big(\partial_{t}+\vec{v}\cdot\vec{\partial}\,\big)\mathsf{P}[{{{E}}, \vec{v}}] = \mathsf{H}[{{{E}}, \vec{v}}]\times \mathsf{P}[{{{E}}, \vec{v}}]\,.
\end{split}
\end{equation}
We use the notation of Ref.~\cite{Bhattacharyya:2020jpj}. The polarization vector $\mathsf{P}[{{{E}}, \vec{v}}]$ for each momentum labeled by $({{{E}}, \vec{v}})$ encodes the flavor state.\footnote{A few words about notation: Functional dependence will be denoted by square brackets $[\ldots]$, though dependence on space-time $(\vec{x},t)$ is implicit. We reserve the parenthesis $(\ldots)$ for grouping terms together or to  denote vectors written as a tuple of their components.  Sans-serif letters such as $\mathsf{S}$ represent three-component vectors in flavor space. Symbols in the usual italic with an arrow on top, e.g., $\vec{x}$, represent three-dimensional  vectors in real space.  Later in the paper we will introduce matrices that are space-time tensors; these are written as bold symbols, e.g., $\mathbf{\Pi}$, and their components as ${\Pi}_{ij}$ with subscripts $ij$ being indices that run over space and time. Angle brackets, i.e., $\langle \cdots \rangle$, will stand for spatial averaging over all spatial coordinates.} Antineutrinos are represented by $\bar{\mathsf{P}}[E,\vec{v}]$. However, polarization vectors for antineutrinos behave as if they were polarization vectors for neutrinos with negative $E$, so it is convenient to define ${\mathsf{P}}[-E,\vec{v}]:=-\bar{\mathsf{P}}[E,\vec{v}]$, where now the argument $E$ in ${\mathsf{P}}[E,\vec{v}]$ takes values between $-\infty$ to $+\infty$. The overall minus sign is a notational foresight that makes Eq.\eqref{eq:hself} simpler.

Neutrino oscillations do not change the total occupation numbers of neutrinos (or antineutrinos), but only the difference between flavors.  One defines ${\mathsf{P}}[E,\vec{v}]=g[{{{E}}, \vec{v}}]\,{\mathsf{S}}[E,\vec{v}]$, with ${\mathsf{S}}[E,\vec{v}]$ having a unit length and
\begin{equation}
g[{{{E}}, \vec{v}}]=\begin{cases}+f_{\nu_e}[{{{+E}}, \vec{v}}]-f_{\nu_\mu}[{{{+E}}, \vec{v}}]& \text{for $E>0$}\\ -f_{\bar\nu_e}[{{{-E}}, \vec{v}}]+f_{\bar\nu_\mu}[{{{-E}}, \vec{v}}]& \text{for $E<0$}\end{cases}\,
\end{equation}
being the normalization of the polarization vectors up to a sign. It is to be noted that the occupation numbers for neutrinos and antineutrinos are only defined at positive $E$, but together they are packaged into a single function $g[E,\vec{v}]$ which spans over $E\in(-\infty,+\infty)$. Typically one has an excess of $\nu_e$  over $\nu_\mu$ (resp. $\bar\nu_e$  over $\bar\nu_\mu$) at any momentum, because the electron flavors can be preferentially produced via charged current processes.  This means that  $g[{{E}, \vec{v}}]$ is positive (resp. negative) for neutrinos (resp. antineutrinos). If instead there is an excess of $\nu_\mu$ over $\nu_e$ (resp. $\bar\nu_\mu$ over $\bar\nu_e$) it simply means that $g[{{E}, \vec{v}}]$ becomes negative (resp. positive). 

In the flavor basis, the orientation $(0,0,+1)$ corresponds to a purely electron flavor and $(0,0,-1)$ to a muon flavor. All neutrinos and antineutrinos start out as flavor eigenstates and the initial state of all Bloch vectors is chosen to be $\mathsf{S}_{\rm ini}[{{E}, \vec{v}}]=(0,0,+1)$. In this convention, at any space-time point, the third component of the Bloch vector $\mathsf{S}[{{E}, \vec{v}}]$ is equal to twice the survival probability minus one.

The Bloch vector for the hamiltonian has the form $\mathsf{H}=\mathsf{H}^{\rm vac}+\mathsf{H}^{\rm mat}+\mathsf{H}^{\rm self}$. Neutrino mass-mixing gives 
\begin{equation}
\mathsf{H}^{\rm vac}[{E}] = \pm\,\omega_{E}  \left(\sin{2\theta}, 0, \cos{2\theta}\right)\,,
\end{equation}
with the plus (resp. minus) sign chosen for the normal (resp. inverted) mass ordering. Note that antineutrinos having been defined as neutrinos with negative $E$ accounts for the sign-flip needed in the mass-mixing hamiltonian. Further, the effect of forward scattering on electrons in the background matter is encoded in
\begin{equation}
\mathsf{H}^{\rm mat} = \lambda \left(0, 0, 1 \right)\,,
\end{equation}
and 
\begin{equation}\label{eq:hself}
\mathsf{H}^{\rm self}[{{\vec v}}] = \sqrt{2} G_{F}\int \frac{d^2\vec{v}\,'}{(2\pi)^{3}}\,\big(1-\vec{v}\cdot\vec{v}{\,'}\big)\int_{-\infty}^{+\infty} E'^2dE'\,g[{{E}{'}, \vec{v}{\,'}}]\,\mathsf{S}[{{E}{'}, \vec{v}{\,'}}]
\end{equation}
is the neutrino-neutrino interaction term that depends on the flavor states of other neutrinos and antineutrinos, and causes collective flavor oscillations. Note that the overall minus sign chosen in the definition the polarization vectors for antineutrinos has converted $\int_{0}^{\infty} E^{2}dE\,(\mathsf{P}-\bar{\mathsf{P}})$ to $\int_{-\infty}^{+\infty} E^{2}dE\,g[{E}, \vec{v}]\,\mathsf{S}[{{E}, \vec{v}}]$.

We will consider astrophysical scenarios where the neutrino density is large, i.e., $\mu \gg \omega_{E}, \lambda$, and neglect the vacuum and the matter terms from the hamiltonian. In this limit, inspection of the above equations shows that the dependence of $\mathsf{S}[{{E}, \vec{v}}]$ on energy drops out.\footnote{Strictly speaking, one must keep $\omega_{E}\neq0$ in $\mathsf{H}^{\rm vac}$; otherwise one can set the mass and flavor basis to be identical, once and for all, and there are no oscillations. Thus $\omega_E$ affects the kickstarting of the oscillations, but the subsequent evolution very weakly~\cite{Chakraborty:2016lct,Dasgupta:2017oko}. In practice, one imagines setting $\omega_E$ (or an external perturbation used as its proxy) to zero immediately after the evolution begins.} So, it makes sense to rewrite Eq.\eqref{1} as
\begin{align}\label{2}
\Big(\partial_{t}+\vec{v}\cdot\vec{\partial}\,\Big)\mathsf{S}[{\vec{v}}] = \mu_{0}\int d^{2}\vec{v}\,'\,
\left(1-\vec{v}\cdot\vec{v}{\,'}\right) G[{\vec{v}'}]\,\mathsf{S}[{{\vec{v}{\,'}}}] \times \mathsf{S}[{{\vec{v}}}]\,,
\end{align}
where we note that $\mathsf{S}$ no longer depends on $E$ but only on  $\vec{v}$, and the collective potential is some constant $\mu_0=\sqrt{2}G_F\,n_\nu$ over length and time scales of interest. We use $\mu_0^{-1}=1$  as our unit of distance.

The function $G[{\vec{v}}]$ is called the electron lepton number (ELN) distribution as a function the direction $\vec{v}$~\cite{Chakraborty:2016lct},
\begin{subequations}
	\begin{align}
G[{\vec{v}}]&=\frac{1}{(2\pi)^3\,n_\nu}\int_0^{\infty}E^2dE\,\Big(f_{\nu_e}[{{{E}}, \vec{v}}] - f_{\bar\nu_e}[{{{E}}, \vec{v}}]-f_{\nu_\mu}[{{{E}}, \vec{v}}]+f_{\bar\nu_\mu}[{{{E}}, \vec{v}}]\Big)\\
&= G_{\nu_e}[{\vec{v}}]-G_{\bar\nu_e}[{\vec{v}}]-G_{\nu_\mu}[{\vec{v}}]+G_{\bar\nu_\mu}[{\vec{v}}]\,.
	\end{align}
\end{subequations}
Here each term on the right hand side is the ratio of the number density of neutrinos or antineutrinos with that flavor and velocity $\vec{v}$ to the neutrino density $n_\nu$. 

\subsection{ELNs with 1,\,2,\,\ldots,\,$\infty$ Crossings}
\label{sec:elntype}

\begin{figure}
\begin{center}
   \hspace{-1.15cm}\includegraphics[width=0.42\columnwidth]{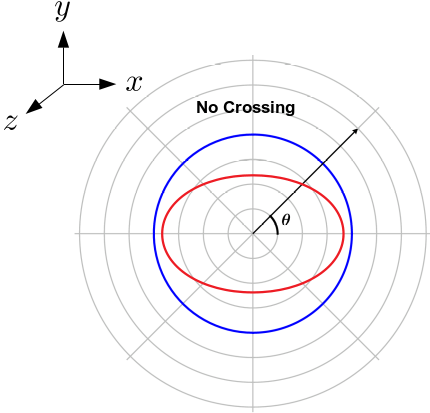} \hspace{0.4 cm} 
	\includegraphics[width=0.35\columnwidth]{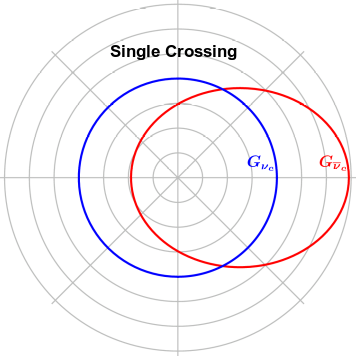}  \\  
	\vspace{0.5 cm}
   \includegraphics[width=0.35\columnwidth]{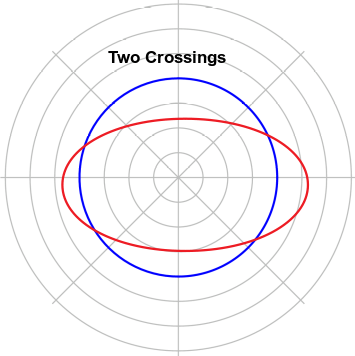} \hspace{0.4 cm}
   \includegraphics[width=0.35\columnwidth]{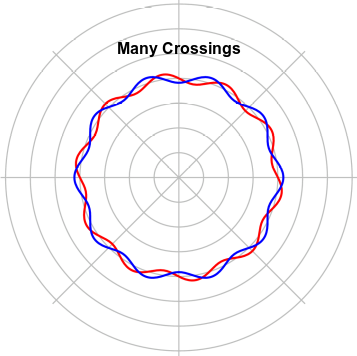} \\
   
\end{center}	
	\caption{Schematic representations of contributions to the ELN, with the four kinds angular distributions of the $\nu_e$ and $\bar\nu_e$ and thus different numbers of crossings.}
	\label{fig1}
\end{figure}

In a typical astrophysical scenario, where the ambient temperature is lower than the muon mass, there is no significant difference in the occupations for $\nu_\mu$ and $\bar\nu_\mu$. Thus $G[{\vec{v}}]$ is approximately equal to $G_{\nu_e}[{\vec{v}}]-G_{\bar\nu_e}[{\vec{v}}]$, and can be called the electron lepton number distribution. Integrating it over velocity, one finds
\begin{equation}
A=\int d^2 \vec{v}\,G[{\vec{v}}] = \frac{n_{\nu_e}-n_{\bar\nu_e}}{n_\nu}\,,
\end{equation}
which is the net lepton asymmetry in neutrinos. 

We now focus on a broad feature of this function $G[\vec{v}]$ --- the possibility that this function changes sign as a function of direction --- a feature that is referred to as an ELN crossing. The reason we focus on this feature, is that it appears to be necessary (and perhaps sufficient) for causing fast flavor oscillations. A crossing can occur if $G_{\nu_e}[{\vec{v}}]$ and $G_{\bar\nu_e}[{\vec{v}}]$ have different velocity dependence and are of a comparable magnitude. Broadly, one can think of four possible scenarios:
\begin{itemize}
\item {{\bf No crossing}: If the density of $\nu_{e}$ far exceeds that of $\bar\nu_{e}$, or vice versa, then one of the terms in $G[\vec{v}]$ dominates and there is no crossing in the ELN. The function $G[\vec{v}]$ remains positive or negative everywhere. This situation is shown as the top-left schematic in Fig.\,\ref{fig1}.}

\item {{\bf One crossing}: In the neutrino decoupling region in a SN, ${\nu}_{e}$ is expected to have a higher number density compared to $\bar\nu_{e}$. However, $\bar\nu_e$ kinematically decouples at a smaller radius, and in the forward direction one may expect the $\bar\nu_{e}$ contribution to the ELN to exceed the $\nu_{e}$~\cite{Sawyer:2015dsa}, if there are directions along which the lepton asymmetry is not too large~\cite{Dasgupta:2016dbv}. In this case, the function $G[\vec{v}]$ is negative in the forward direction, but positive elsewhere, and there is a single crossing. Hydrodynamic fluctuations of the lepton asymmetry can allow asymmetry to be small along some direction(s) and ELN crossing(s) could occur~\cite{Morinaga:2019wsv,DelfanAzari:2019tez,Abbar:2019zoq}. As another example, in the disk of a neutron star merger~\cite{Wu:2017qpc,George:2020veu} (or if a quark-hadron phase transition occurs in a SN~\cite{Dasgupta:2009yj}) the number density of ${\bar\nu}_{e}$ can exceed that of $\nu_{e}$, and all else remaining the same, a $\bar\nu_e$ excess in the radially outward direction is even more likely. This situation is shown as the top-right schematic in Fig.\,\ref{fig1}.}

\item {{\bf Two crossings}: If the $\bar\nu_{e}$ contributions to the ELN exceed the $\nu_{e}$ in the forward and backward directions, the function $G[\vec{v}]$ is negative in the forward and backward directions, but positive in the directions tangential to the radial direction. A possible cause of larger $\bar\nu_e$ numbers in the backward direction, in addition to a forward excess as discussed above, could be that $\bar\nu_e$ have a larger cross section to backscatter off nuclei, thus creating a dominantly $\bar\nu_e$ back-flux~\cite{Morinaga:2019wsv}. This situation with two crossings is shown as the bottom-left schematic in Fig.\,\ref{fig1}.}

\item {{\bf Many crossings}: If the $\nu_{e}$ and $\bar\nu_{e}$ contributions to the ELN are almost equal but fluctuate independently, perhaps due hydrodynamic waves and instabilities, one would expect $G[\vec{v}]$ to be changing sign frequently as a function of direction. In this scenario, the ELN has many crossings. This is likely in the convective layer of a SN~\cite{DelfanAzari:2019tez,Abbar:2019zoq,Glas:2019ijo}, and perhaps in the early Universe~\cite{Hansen:2020vgm}. A schematic of this situation is shown on the bottom-right in Fig.\,\ref{fig1}.}

\end{itemize} 

\subsubsection{True vs. Apparent Dimensionality}
\label{sec:truevsapp}

The figures in the schematic in Fig.\,\ref{fig1} are best visualized as sections of the corresponding surfaces in the three-dimensional velocity space. However, studying fast flavor oscillation in three spatial dimensions is numerically challenging.
Typically, one assumes that $G[{\vec{v}}]$ is azimuthally symmetric, i.e., invariant to its rotations in velocity-space about the radially outward direction (here coinciding with $\hat{x}$). This is the picture that has been adopted in almost all studies. Mathematically, this requires dropping $v_{z}$ and $v_{y}$ everywhere and setting $\vec{v}\cdot\vec{\partial} \to v_{x}\partial_{x}$ and $\vec{v}\cdot\vec{v}\,'\to v_{x}\,v'_{x}$ in Eq.\,\eqref{2}. Thus, in most studies, even when one solves the counterpart of Eq.\eqref{1} in some lower number of spatial dimensions, say $d=1$ assuming azimuthal symmetry about the radially outward direction, what one has in mind is the higher dimensional version, say with $D=3$, and the solution is assumed to be strictly symmetric with respect to the remaining $D-d=2$ components of the velocity, $v_{y}$ and $v_{z}$. In this approach ${v}_{x}$ does not have a unit magnitude, and one only requires $|v_{x}| < 1$. Although it is prevalent and appealing as a modeling simplification, it is an uncontrolled approximation. In previous studies on slow collective oscillations~\cite{Raffelt:2013rqa, Mangano:2014zda,Dasgupta:2015iia,Abbar:2015fwa}, it was shown that additional instabilities can be generated by spontaneous symmetry breaking along the assumed-to-be-symmetric dimensions.

In this paper we will consider a neutrino gas in strictly two spatial dimensions, $x$ and $y$, with velocities restricted to a two-dimensional plane, i.e., $\vec{v}=v_{x}\hat{x}+v_{y}\hat{y}$ with $v_{x}^{2}+v_y^2 = 1$. This should \emph{not} be thought of as a projection of a three dimensional problem on to two dimensions, as described in the paragraph above. In this way, the velocity vector has a unit length and no instabilities are ignored by fiat. The downside is that we are solving a two-dimensional toy problem, and it may not be obvious how it applies to the three-dimensional real-world. Although our calculation in 2+2+1 dimensions is the state-of-the-art, it should really be taken as a step towards more realistic studies in full 3+3+1 dimensions. Still, this set-up has its merits as discussed above, and we will find important insights by undertaking this calculation.

\subsubsection{ELN Models}

In this strictly two-dimensional approach, the ELNs in the figure are functions of a single independent variable, say $\theta$, and one writes $G[\vec{v}]=G[\theta]$, where $v_x = \cos{\theta}$ and $v_y = \sin{\theta}$. In the remainder of our study, we consider three out of the four zero crossing scenarios shown in Fig.\ref{fig1}. We ignore the case with no crossings because one finds no instabilities in that case.\footnote{One point of semantics before we proceed further: For a three-dimensional but azimuth-symmetric ELNs these correspond to \emph{one closed curve}, \emph{two closed curves}, and \emph{many closed curves} worth of crossings. For our strictly two-dimensional ELNs we will continue to call the above ELNs as having one, two, or many crossings, though strictly speaking they have two, four, and twice as many crossings, respectively.} 
We further simplify our ELNs to be piecewise constant or sinusoidal, for concreteness, and consider 
\begin{itemize}

\item Type\,I (single crossing): $G[\theta]=\begin{cases}
\frac{A-1}{2\pi}, &{\rm if} \, v_{x}=\cos\theta>0\\
\frac{1}{2\pi}, &{\rm if} \, v_{x}=\cos\theta<0\,.
\end{cases}$

\item Type\,II (two crossings): $G[\theta]=\begin{cases}
\frac{A-1}{2\pi}, &{\rm if} \, v_{x}=\cos\theta>0 ~\mbox{{\&}}~ v_{y}=\sin\theta>0\\
\frac{1}{2\pi}, &{\rm if} \, v_{x}=\cos\theta<0 ~\mbox{{\&}}~ v_{y}=\sin\theta>0\\
\frac{A-1}{2\pi}, &{\rm if} \, v_{x}=\cos\theta<0 ~\mbox{{\&}}~ v_{y}=\sin\theta<0\\
\frac{1}{2\pi}, &{\rm if} \, v_{x}=\cos\theta>0 ~\mbox{{\&}}~ v_{y}=\sin\theta<0\,.
\end{cases}$

\item Type\,III (many crossings): $G[\theta]= \frac{A}{2\pi}+c_{1}\,\cos m\theta + c_{2}\,\sin m\theta$\,.

\end{itemize}

The above choices are made in a way such that $G[\theta]$ remains unchanged under $v_y \rightarrow -v_y$ for Type\,I, and a simultaneous exchange $v_y \rightarrow -v_y$ and $v_x \rightarrow -v_x$ in case of Type\,II. These choices are made to explore the dependence of the solution on the nature of symmetry of $G[\theta]$. For Type\,III, $m$ will be taken to be large and $G[\theta]$ can have ${\cal O}(m)$ number of zero crossings as a function of $\theta$. In all these cases, 
\begin{equation}
A = \int_{0}^{2\pi} d\theta\,G[\theta]
\end{equation}
denotes the lepton asymmetry.


\subsection{Linear Stability Analysis}
\label{sec:lsa}
Initially the transverse component of the Bloch vector is small, i.e., \mbox{$\mathsf{S}^{\perp}[{v_x, v_y}]\ll 1$}~\cite{Banerjee:2011fj}, as the neutrinos start out as flavor pure states. We then write its space-time evolution to linear order as~\cite{Chakraborty:2016lct,Dasgupta:2016dbv}
\begin{equation}\label{3}
\begin{split}
\left(\partial_{t}+v_x\partial_x + v_y \partial_y\right)\mathsf{S}^{\perp}[{v_x, v_y}] &= i\int_{-1}^{+1} \int_{-1}^{+1} dv'_x dv'_y\,\delta[v{'}-1] \left(1-v_xv_x{'}-v_yv'_y\right) {G}[{v'_x, v'_y}] \\
&\times\left(\mathsf{S}^{\perp}[{v'_x, v'_y}]-\mathsf{S}^{\perp}[{v_x, v_y}]\right)\,.
\end{split}
\end{equation}
where $v = \sqrt{v_x^2 + v_y^2}$. This is a linear equation in $\mathsf{S}^{\perp}$. It is thus natural to decompose it in the Fourier basis, 
\begin{equation}\label{4}
\mathsf{S}^{\perp}[{v_x, v_y}] = \sum_{\vec{K},\Omega}\mathsf{Q}_{\vec{K},\Omega}^{\perp}[{v_x, v_y}]e^{i (K_{x}x+K_{y}y-\Omega t)}\,, 
\end{equation}
which gives a linear algebraic equation connecting $\Omega$ with $\vec{K}$. Once this relationship $\Omega(\vec{K})$ is known, it gives a set of basis functions $\mathsf{Q}$ labelled by $(\vec{K},\Omega)$ which can be linearly superposed to describe any solution of $\mathsf{S}^{\perp}[{v_x, v_y}]$ in the linear regime. 

Concretely, one finds the dispersion relation~\cite{Chakraborty:2016lct,Dasgupta:2016dbv,Izaguirre:2016gsx,Capozzi:2017gqd,Morinaga:2018aug,Airen:2018nvp,Yi:2019hrp,Chakraborty:2019wxe, Doring:2019axc}
\begin{equation}\label{5}
\mathcal{D}
= det\,\mathbf{\Pi}[k_{x}, k_{y}, \omega] = 0\,,
\end{equation}
where 
\begin{equation}\label{6}
\mathbf{\Pi} [k_{x}, k_{y}, \omega] = \boldsymbol{\eta} + \int_{-1}^{1}\int_{-1}^{1} dv_{x} dv_{y} \hspace{1mm}\frac{G[v_{x}, v_{y}]}{\omega - v_xk_x-v_yk_y}\,\delta[v-1]\, \mathbf{W} [v_x, v_y]\,.
\end{equation}
If there is a solution that grows exponentially, e.g., ${\rm Im}\,\omega>0$, that solution is said to be unstable~\cite{Capozzi:2017gqd}. A more detailed classification can be found in Refs.~\cite{Airen:2018nvp,Yi:2019hrp}. In Eq.\eqref{6} the following definitions have been used: 
\begin{align}\label{7}
\boldsymbol{\eta} = \begin{pmatrix}
1 & \phantom{-}0 & \phantom{-}0 \\
0 & -1 & \phantom{-}0 \\
0& \phantom{-}0& -1 
\end{pmatrix}\,,
\end{align}
and 
\begin{align}\label{8}
\mathbf{W}[v_x, v_y] = \begin{pmatrix}
1 & v_x & v_y \\
v_x & v_{x}^2 & v_{x}v_{y} \\
v_{y} &  v_{x}v_{y} & v_{y}^2 
\end{pmatrix}\,,
\end{align}
and
\begin{subequations}
	\begin{align}
	\omega &= \Omega - \phi_{tt}\,, \label{9a}\\
    k_x &= K_x - \phi_{tx} \,, \label{9b}\\
	k_y &= K_y - \phi_{ty}\,, \label{9c}
	\end{align}
\end{subequations}
wherein 
\begin{align}\label{10}
\boldsymbol{\phi} = \int_{-1}^{1}\int_{-1}^{1} dv_{x} dv_{y} \hspace{1mm}G[{v_{x}, v_{y}}]\,\delta[v-1]\,\mathbf{W} [v_x, v_y]\,.
\end{align}

Thus, in the linear regime, allowed solutions are given by Eq.\eqref{5}, which upon after expanding gives
\begin{equation}\label{11}
\begin{split}
-\left(\Pi_{ty}\right)^2 \Pi_{xx} + 2 \Pi_{tx}\Pi_{ty}\Pi_{xy} -  \Pi_{tt} \left(\Pi_{xy}\right)^2 - \left(\Pi_{tx}\right)^2 \Pi_{yy} + \Pi_{tt} \Pi_{xx} \Pi_{yy} 
= 0
\end{split}
\end{equation}
For each pair $(k_{x}, k_{y})$ one needs to solve Eq.\eqref{11} for $\omega$, whose imaginary part describes the growth of the flavor instabilities. Note that because the only dimensionful quantity in the problem is $\mu_0$, these are all fast instabilities, i.e., ${\rm Im}\,\omega \propto \mu_0$.
In general, the above equation is transcendental and analytical solution is impossible. However, for $(k_{x} = 0, k_{y} = 0)$, Eq.\eqref{11} becomes a simple cubic equation in $\omega_{0} = \omega[k_{x}=0, k_{y} = 0]$,
\begin{equation}\label{12}
\omega_{0}^{3}+\gamma_2 \omega_{0}^{2}+\gamma_1 \omega_{0} + \gamma_0 = 0\,,
\end{equation}
where
\begin{subequations}
\begin{align}
    \gamma_2 &= \phi_{tt}-\phi_{xx}-\phi_{yy}\,, \label{13a}\\
    \gamma_1 &= \left(\phi_{tx}\right)^{2} + \left(\phi_{ty}\right)^{2}-\phi_{tt} \phi_{xx} - \left(\phi_{xy}\right)^{2} - \phi_{tt} \phi_{yy} + \phi_{xx} \phi_{yy}\,, \label{13b}\\
    \gamma_0 &= -\left(\phi_{ty}\right)^2 \phi_{xx} + 2 \phi_{tx} \phi_{ty} \phi_{xy} - \phi_{tt} \left(\phi_{xy}\right)^2 -\left(\phi_{tx}\right)^{2} \phi_{yy} + \phi_{tt} \phi_{xx} \phi_{yy}\,. \label{13c}
\end{align}
\end{subequations}
We remind the reader that the $\omega$ here is \emph{not} $|\Delta m^{2}|/(2E)$, but merely the zeroth component of the Fourier mode in Eq.\eqref{9a}. The values of $\left(\gamma_0, \gamma_1, \gamma_2 \right)$ and $\phi_{ij}$ for our chosen Type I, II, and III neutrino angular distributions are listed in Table\,\ref{table:Table}.
\begin{table}[t]\caption{Elements of $\boldsymbol\phi$ for various types of ELNs}
	\centering
	\renewcommand{\arraystretch}{1.6}
	\begin{tabular}{c c c c c c c c c c }
    \hline
    ELN  & $\phi_{tt}$ & $\phi_{tx}$ & $\phi_{ty}$ & $\phi_{xx}$ & $\phi_{yy}$ & $\phi_{xy}$ & $\gamma_2$ & $\gamma_1$ & $\gamma_0$\\ [0.7ex]
    \hline
    Type\,I & $\frac{A}{\pi}$ & $\frac{\left(2-A \right)}{4}$ & 0  & $\frac{2A}{3 \pi}$ & $\frac{A}{3 \pi}$ & 0 & 0 & -$\frac{7 A^2}{9 \pi^2}$+$\frac{\left(A-2\right)^2 }{16}$ &  $\frac{2 A^3}{9\pi^3}$-$\frac{A\left(A-2\right)^2 }{48 \pi}$  \\
    Type\,II & $\frac{A}{\pi}$ & 0 & 0  & $\frac{2A}{3\pi}$ & $\frac{A}{3\pi}$ & $\frac{A-2}{3\pi}$ & 0 & $\frac{-8A^2+4A-4}{9 \pi^2}$& $\frac{A(A^2+4A -4)}{9\pi^3}$\\
    Type\,III & A & 0 & 0  & $\frac{A}{2}$ & $\frac{A}{2}$ & 0 & 0 & $\frac{-3 A^2}{4}$ & $\frac{A^3}{4}$\\
    \hline
    \end{tabular}
\label{table:Table}
\end{table}
\subsection{EoM Solver}
\label{sec:numstrat}
We developed our own numerical routines for solving Eq.\eqref{2}. Our approach involves discretizing the spatial directions into $N_{x}$ and $N_{y}$ uniformly spaced bins, resulting in $N_{x}\,N_{y}$ number of coupled nonlinear ODEs in time for each momentum mode labeled by its $(v_{x}, v_{y})$ pair. A periodic boundary condition in each spatial direction is assumed. We also discretize the velocity modes in one direction (either $v_{x} \hspace{1mm}\rm{or}\hspace{1mm}$$v_{y}$) into $N_{vel}$ uniformly spaced bins, and due to the restriction $v = 1$ the binning of the velocity modes in the other direction gets fixed resulting in total of $\left(3\times 2\right)\,N_{x}\,N_{y}\,N_{vel}$ coupled nonlinear ODEs. The factor of 3 comes from the three components of the polarization vector and the factor of 2 comes from the fact that for each choice of $v_x$ one has two allowed choices of $v_y$.

We solve the system equations in a 2D square box of area $L \times L$,  with $L = 18$ in units of $\mu_0^{-1}$. We choose $ \mu_0 = 3 \pi\, \rm{cm^{-1}}$ for Type\,I ELNs, and for Type\,II ELNs we take either $ \mu_0 = 3 \pi\,\rm{cm^{-1}}$ or $\mu_0 = 17\pi\,\rm{cm^{-1}}$ which correspond to a neutrino number density of ${\cal O}\left(10^{32}\right) \, \rm{cm^{-3}}$. Periodic boundary conditions are assumed on both spatial directions, i.e., on $x, y \in \left(-\frac{L}{2}, \frac{L}{2}\right)$. This periodic boundary condition physically represents that we are treating this box as a part of a larger system. The finiteness of the box affects the smallest $\Delta\vec{k}$ we can distinguish in our calculation. We discretize $x$ and $y$ into $N_x = N_y = 480$ bins, enough to trigger as many Fourier modes as possible, and well above what is needed to trigger all unstable $\vec{k}$ modes, limited only by CPU hours. The velocity of outgoing neutrinos are in the range $v_x, v_y \in \left(-1, 1 \right)$ with $N_{vel} = 32$. In total, we solve a system of $6 N_x \times N_y \times N_{vel} = 44236800$ coupled nonlinear ODEs in time up to $t_{\rm fin} = 3.5$ in units of $\mu_0^{-1}$. The choices for $N_{x}, N_{y}, N_{vel}$ are optimized to obtain sufficient precision and accuracy as shown in Appendix\,\ref{sec:appendix}.

The initial conditions are that all Bloch vectors $\mathsf{S}[{\vec{v}}]$ are equal to $(0,0,1)$, i.e., a flavor-pure state. Depending on the positive (or negative) sign of $G[{\vec{v}}]$, the polarization vectors $\mathsf{P}[{\vec{v}}]=G[{\vec{v}}]\,\mathsf{S}[{\vec{v}}]$ points along (or opposite to) the vertical in flavor space. For our chosen set of ELNs, as discussed in Sec.\,\ref{sec:elntype}, this means that the $\mathsf{P}[{\vec{v}}]$ start with one, two, or many crossings\footnote{See the footnote in Sec.\,\ref{sec:elntype}, clarifying what we mean by one, two, and many crossings in the context of our ELNs.} as a function of $\theta$. Normally $\mathsf{H}_{\omega}^{\rm{vac}}$ would start the flavor evolution by tilting the Bloch vectors away from their initial positions. However, for our calculations, we have set $\mathsf{H}_{\omega}^{\rm{vac}}$ and $\mathsf{H}^{\rm{mat}}$ to zero for numerical convenience, and instead provide an external perturbation of ${\cal O}(10^{-6})$ to both transverse components of the polarization vectors at $(x=0,y=0)$ to kickstart the evolution. 

The code is written in {\sc Python} and uses the {\sc zvode} solver, a variable-coefficient differential equation solver in {\sc Python}, to solve the system of ODE as a function of time. This solver implements the backward differentiation formula for numerical integration. Our technique of converting a set of coupled nonlinear PDEs into ODEs allows easy use of existing ODE libraries and makes the numerical integration much faster. The spatial derivatives are computed using a Fast Fourier Transform employing {\sc Python}'s {\sc scipy.fftpack.diff} package.

\subsection{Dispersion Relation Solver}
\label{sec:anastrat}

To understand the nature of the flavor evolution in the linear regime one needs to know the behavior of $\omega$ as a function of $\vec{k}$, which we represent by its magnitude $k$ and argument as $\beta$, i.e., 
\begin{eqnarray}
k &= \sqrt{k_x^2+k_y^2}\,,\\
\beta &= \arctan \left(\frac{k_y}{k_x}\right)\,.
\end{eqnarray}
This $\omega[k,\beta]$ can only be obtained after solving the transcendental equation described by Eq.\eqref{11}. This is not easy, even numerically, as one needs to scan over a large space spanned by $(\omega,k,\beta)$, and brute force root-finding is inefficient.


To speed up our root-finding, we use the method of iterative solving where we begin at a known solution (or initial guess) and then use it to propagate the solution further in $(\omega,k,\beta)$ space.
As our initial guess we use the $k = 0$ solution, $\omega_{0}$, which for a given ELN can be determined easily from Eq.\eqref{12}. This is simply the zero-mode solution that was advocated in Ref.\cite{Dasgupta:2018ulw}, and even in the most general case with 3+3+1 dimensions Eq.\eqref{12} is analytically tractable. Then we define a circular boundary of radius $r$, chosen to be sufficiently small and close to $k = 0$ point in the $k-\beta$ plane. Using $w_{0}$ as our initial guess we numerically solve Eq.\eqref{11} to calculate $\omega$ for different $\beta$ directions within the region $0 < k <r$ in the $k-\beta$ plane. Then we proceed to a new point on the boundary defined by $r$, where we already have a solution, to define another circle of radius $r$. At each new point within this new circular region we can start with previous solution as a starting guess, and find the updated solutions. Note that we choose $r$ in a way such that the previous guess works reasonably well. Repeating this, we can find the solution on the entire $k-\beta$ plane.

The roots of Eq.\eqref{11} are obtained by  {\sc Python's} {\sc fsolve} package which uses Powell's conjugate direction method to find the local minima of a nonlinear equation. The method requires an initial guess, but does not require differentiability of the underlying complex function because no derivatives are computed in order to find the solution. The integrals in Eq.\eqref{6} are evaluated using the numerical routine for adaptive quadrature implemented in {\sc Python's} {\sc quad} solver.

\section{Results}
\label{sec:res}

\subsection{One Crossing}
\label{sec:single}
 \begin{figure}
\begin{center}
	\textbf{Type\,I, A = 0}\par\medskip
	\includegraphics[width=0.5\columnwidth]{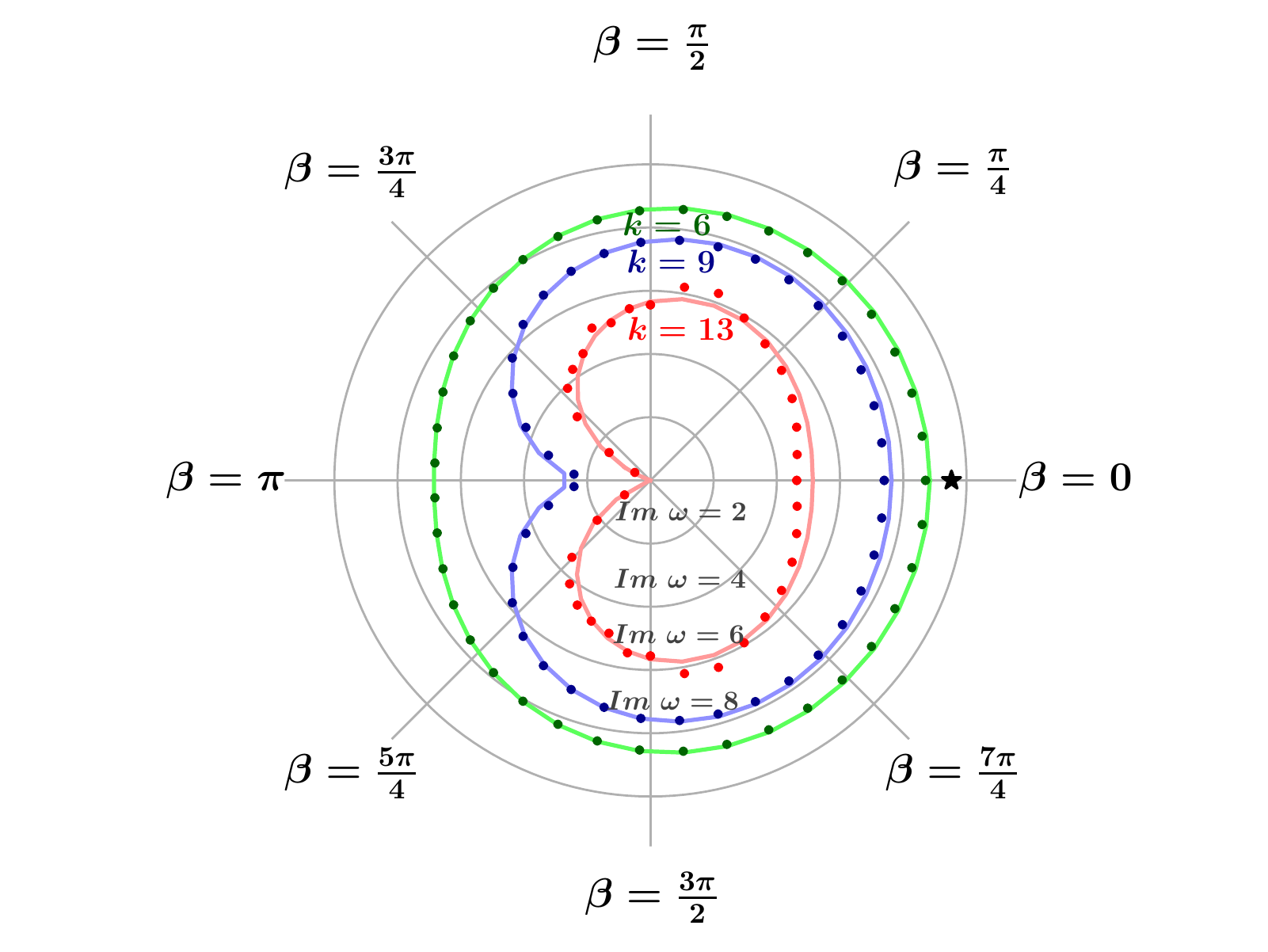}~
	\hspace{0.225 cm}
	\includegraphics[width=0.35\columnwidth]{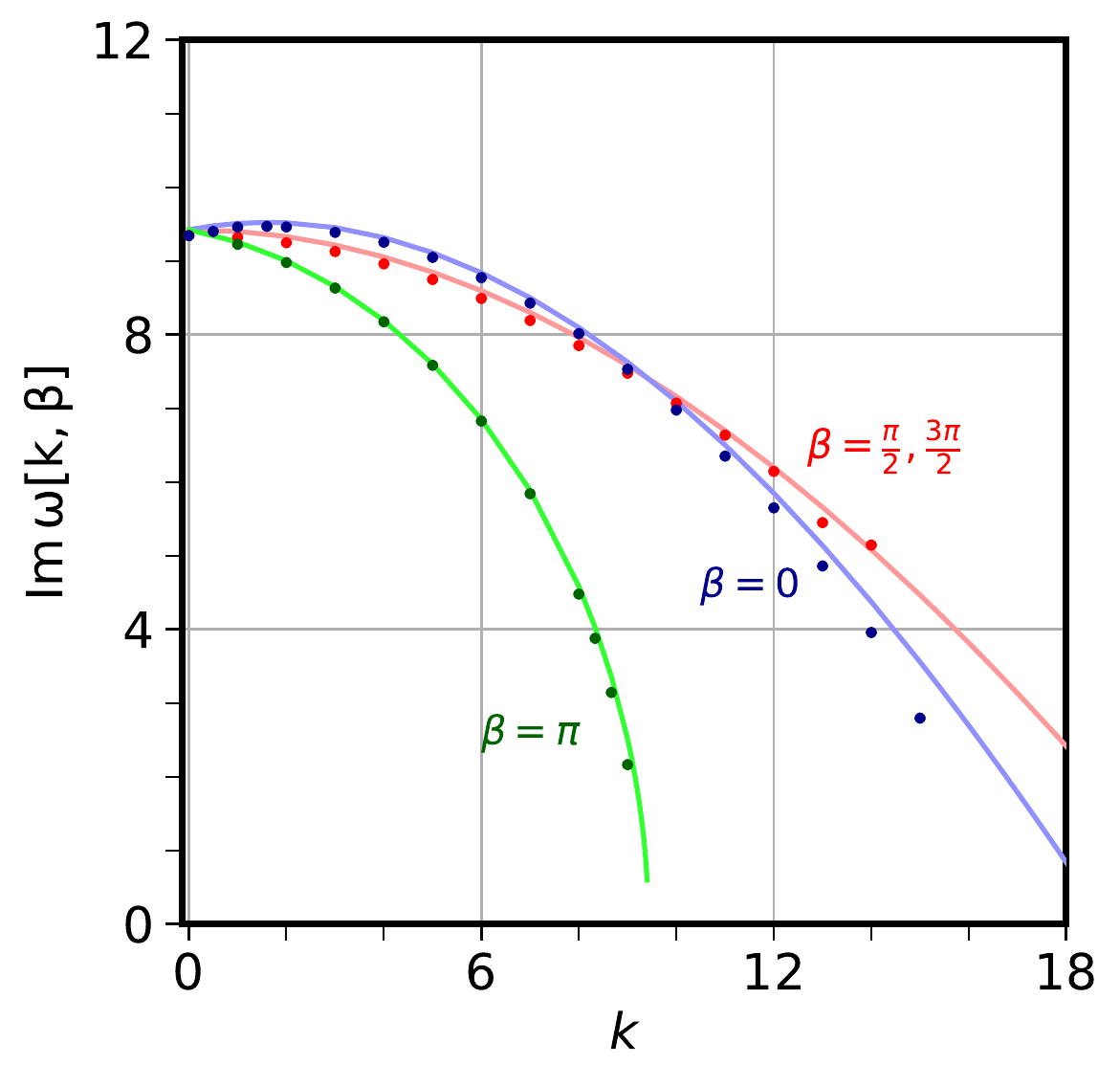}\\[1em]
     \textbf{Type\,I, A $\neq$ 0}\par\medskip
     \includegraphics[width=0.5\columnwidth]{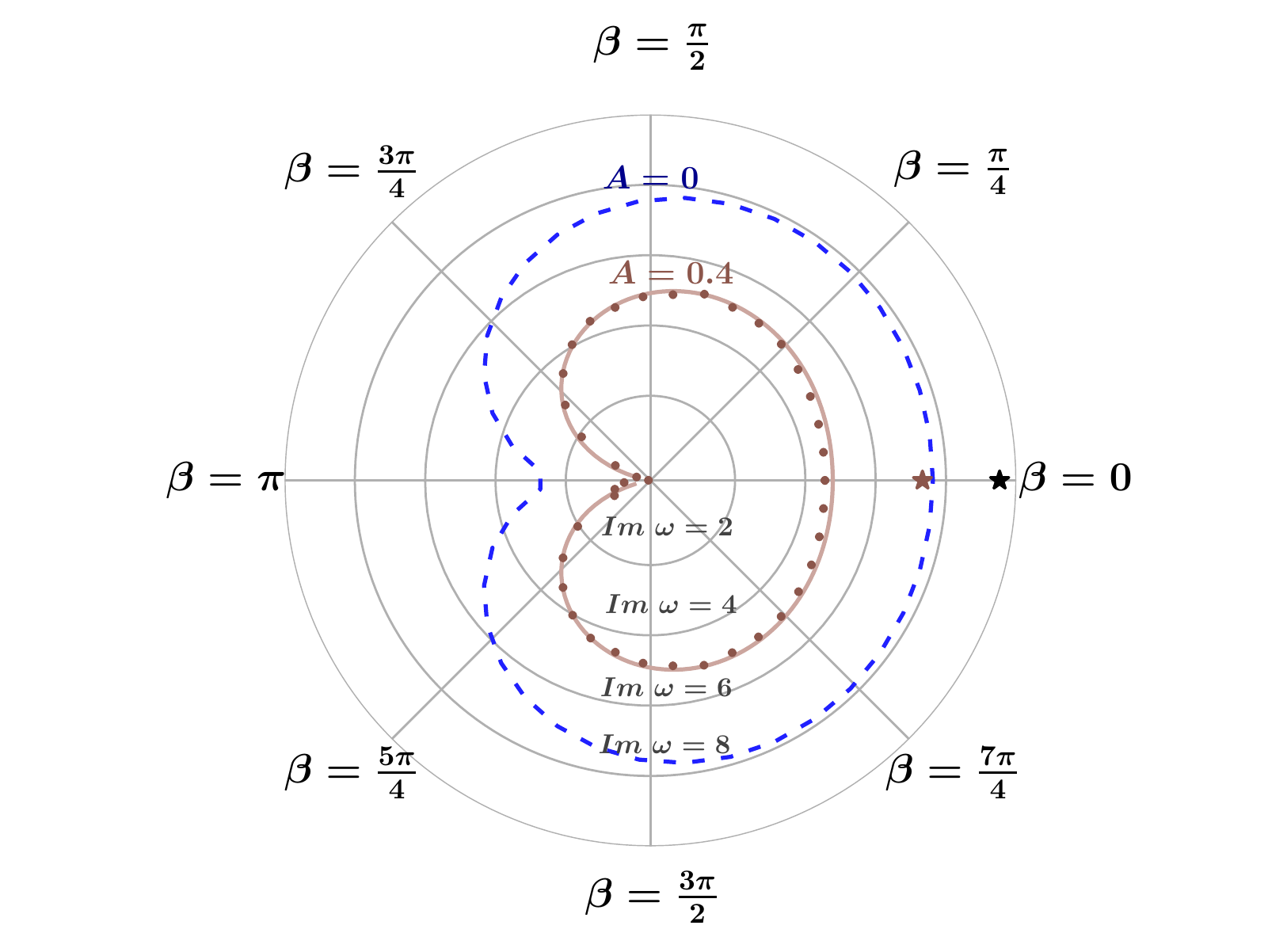}~
      \hspace{0.225 cm}
      \includegraphics[width=0.35\columnwidth]{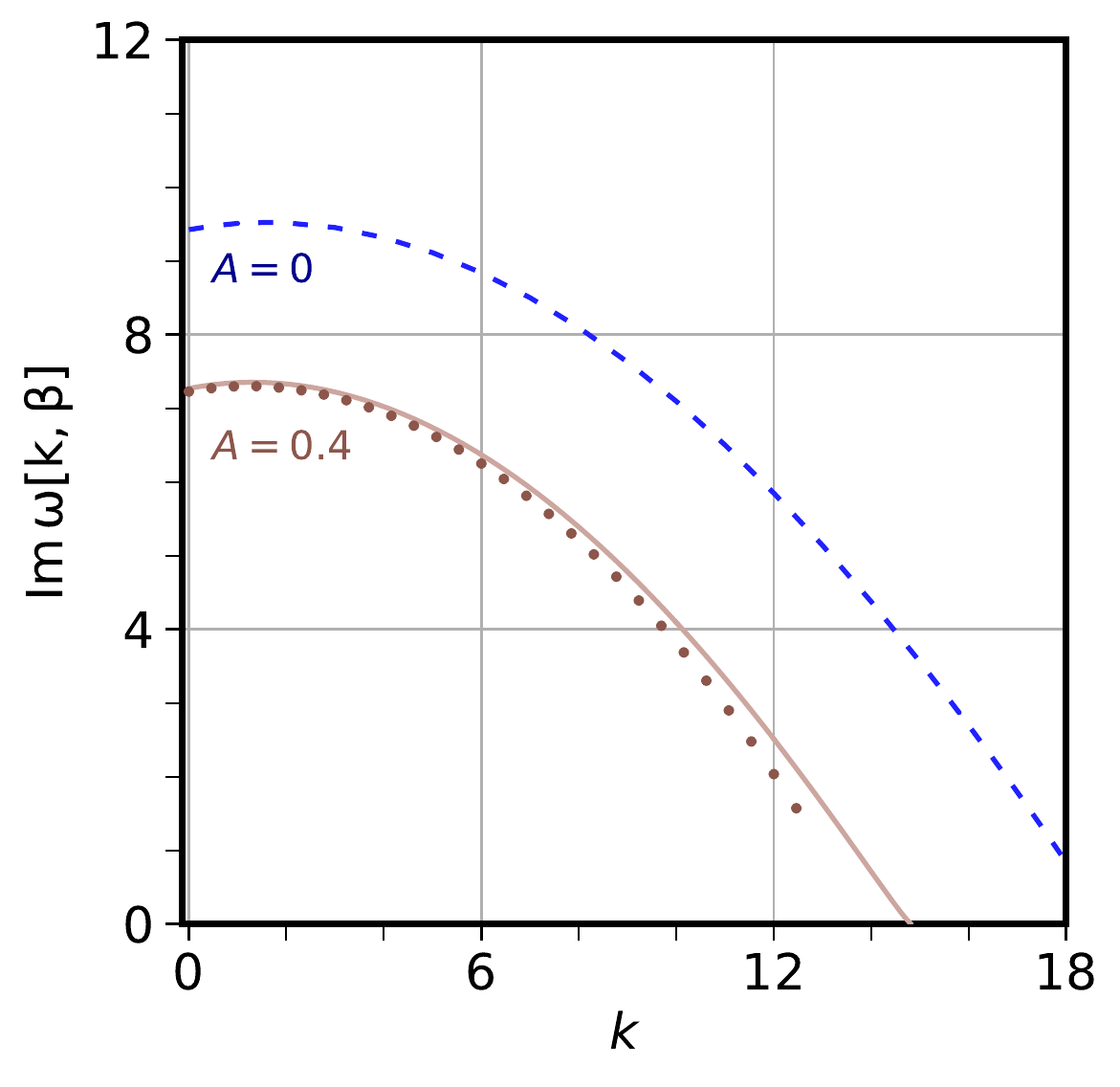}
\end{center}      
\caption{Left: Angular variation of $\rm{Im} \, \omega$ with respect to $\beta$. Right : Radial variation of $\rm{Im} \, \omega$ with respect to $k$. The top plots are done with $A = 0$ for different values of $k$ (left) and $\beta$ (right) whereas the bottom ones with fixed $k = 9$ (left) and $\beta =0$ (right) for $A = 0.4$ case. The continuous lines show results of the linear stability analysis, while the dotted points show the results from numerical solution of the full equations of motion. For comparison, the results for the $A = 0$ case are shown in blue dashed lines in the bottom panel plots.}	
	\label{fig3}
\end{figure}
For our Type\,I ELNs, $G[\theta]$ has a reflection symmetry for $v_y = \sin \theta$, i.e., $G[\theta]$ remains invariant under $v_y \rightarrow - v_y$. This symmetry leads to a similar symmetry in the $k_x-k_y$ plane, i.e., $\mathcal{D}[k_{x}, -k_{y}, \omega] = \mathcal{D}[k_{x}, k_{y}, \omega]$. This can be understood considering the interchange $v_y \rightarrow -v_y$ and $k_y \rightarrow -k_y$ in Eq.\eqref{6}:
\begin{equation}\label{14}
\begin{split}
{\Pi}_{ij}[k_{x}, k_{y}, \omega] &\xrightarrow {\text{$k_y \rightarrow -k_y$}} {\eta}_{ij} + \int_{-1}^{1}\int_{-1}^{1} dv_{x} dv_{y} \hspace{1mm}\frac{G[{v_{x}, v_{y}}]}{\omega - k_xv_x+k_yv_y}\,\delta[v-1]\,{W}_{ij}[v_x, v_y]\\ &\xrightarrow {\text{$v_y \rightarrow -v_y$}} \pm   {\Pi}_{ij}[k_{x}, k_{y}, \omega]\,,
\end{split}
\end{equation}
where we have used the fact that $G[{v_x, v_y}] = G[{v_x, -v_y}]$ in the last step. Eq.\eqref{14} basically says that ${\Pi}_{ij}$ remains invariant under the above two operations up to a $\pm$ sign. The minus sign occurs only for $\Pi_{ty}$ and $\Pi_{xy}$ while all others come with a plus sign. However, $\Pi_{ty}$ and $\Pi_{xy}$ always come in pairs in  Eq.\eqref{11}, i.e., as $\left(\Pi_{ty}\right)^2, \left(\Pi_{xy}\right)^2$ or $\Pi_{ty}\Pi_{xy}$, which immediately says that $\mathcal{D}[k_{x}, k_{y}, \omega]$, as well as the solution for $\rm{Im} \, \omega$, remains invariant under $k_y \rightarrow -k_y$.

Eq.\eqref{14} implies for $k = 0$ or $\left(k_x = k_y = 0\right)$ mode:
\begin{equation}\label{15}
\Pi_{ty}[0, 0, \omega_{0}] = 0
\end{equation}
and 
\begin{equation}\label{16}
\Pi_{xy}[0, 0, \omega_{0}] = 0\,.
\end{equation}
Eq.\eqref{15} and Eq.\eqref{16} help us to write the full dispersion relation in Eq.\eqref{11} as two separate equations:
\begin{equation}\label{17}
\Pi_{yy}[0, 0, \omega_{0}] = 0 
\end{equation}
and 
\begin{equation}\label{18}
\begin{split}
\Pi_{tx}[0, 0, \omega_{0}]\Pi_{tx}[0, 0, \omega_{0}] - \Pi_{tt}[0, 0, \omega_{0}]\Pi_{xx}[0, 0, \omega_{0}]= 0\,.
\end{split}
\end{equation}{\normalsize }
Eq.\eqref{17} implies $\omega_{0} = \phi_{yy}$ which is a real solution. 
Eq.\eqref{18} can be simplified to obtain a quadratic equation in $\omega_0$ as,
\begin{equation}\label{19}
\omega_{0}^{2}+\omega_{0} \Bigl(\phi_{tt} - \phi_{xx}\Bigr) - \Bigl(\phi_{tt}\phi_{xx}-\left(\phi_{tx}\right)^{2}\Bigr) = 0
\end{equation}
The solutions determined by Eq.\eqref{19} can be complex only if
\begin{equation}\label{20}
\bigl(\phi_{tt} + \phi_{xx}\bigr)^{2}-4\left(\phi_{tx}\right)^{2} < 0\,. 
\end{equation}
Eq.\eqref{19} can have complex solutions, in general, leading to the $k = 0$ mode becoming unstable for Type\,I cases. For instance, in our numerical examples with $A = 0$ (resp. $A=0.4$) the LHS of Eq.\eqref{20} becomes $-1$ (resp. $-0.6$), and thus easily satisfies the above condition. Once we have the $k=0$ solution, $\omega_{0}$, we can compute $\omega$ for other value of $\vec{k}$ using the iterative method described in Sec.\,\ref{sec:anastrat}. 
On the other hand, we can also numerically simulate Eq.\,\eqref{2} to obtain $\mathsf{S}[{\vec{v}}]$ as a function of $(x,\,y,\,t)$. We then take spatial Fourier transforms of this solution, and ask how the amplitude of each $\vec{k}$ mode changes with time. For some modes, we find the mode-amplitude increases exponentially, and we extract the imaginary part of $\omega(\vec{k})$ from numerical data. These two methods give results in excellent agreement, as shown in Fig.\,\ref{fig3}.

For Type\,I ELNs, $G[\theta]$ is symmetric between the regions $\sin \theta > 0$ and $\sin \theta < 0$ or in other words there is a $v_y \rightarrow -v_y $ symmetry. This gives rise to a similar symmetry in the angular variation of $\rm{Im} \, \omega$ with respect to $\beta$, i.e., $\rm{Im} \, \omega$$|_{k_y >0}$ = $\rm{Im} \, \omega$$|_{k_y <0}$ as can be seen in the top left panel plot of Fig.\ref{fig3}. The radial variation of $\rm{Im} \, \omega$ with respect to $k$ in the top right plot of Fig.\ref{fig3} clearly shows a much larger growth for the modes very close to $k = 0$, and then the growth rate decreases as a function of $k$ for all $\beta$ directions. The $k = 0$ mode being unstable for this case is understood from our analytical arguments. We find the decrease is much slower along the  $k_y$ axis, about which there is a $k_y \rightarrow -k_y$ symmetry. Interestingly the position of the Fourier mode with the maximum linear growth rate is aligned along $k_x$ axis, as shown in the black starred point in top left plot of Fig.\,\ref{fig3}.

Even for $A\neq0$ a similar kind of $k_y \rightarrow -k_y$ symmetry in the angular variation of $\rm{Im}\, \omega$ is shown in the bottom left plot of Fig.\,\ref{fig3}. The $k = 0$ mode is unstable for this case as well. All the results for the radial and angular variation of $\rm{Im}\, \omega$ are similar to the $A = 0$ case with only an exception that the overall growth rate as well as the maximum growth rate decreases for larger (positive) $A$. This is seen in the bottom left panel of Fig.\,\ref{fig3}, where the position of the maximum growth rate is indicated by different starred points that correspond to specific choices of $A$, as indicated by the color code. This effect of non-zero lepton asymmetry also results in a faster decrease of $\rm{Im}\, \omega$ with larger $k$, as shown in bottom right plot of Fig.\,\ref{fig3}.

\subsection{Two Crossings}
\label{sec:double}
\begin{figure}
\begin{center}
	\textbf{Type\,II, A = 0}\par\medskip
	\includegraphics[width=0.5\columnwidth]{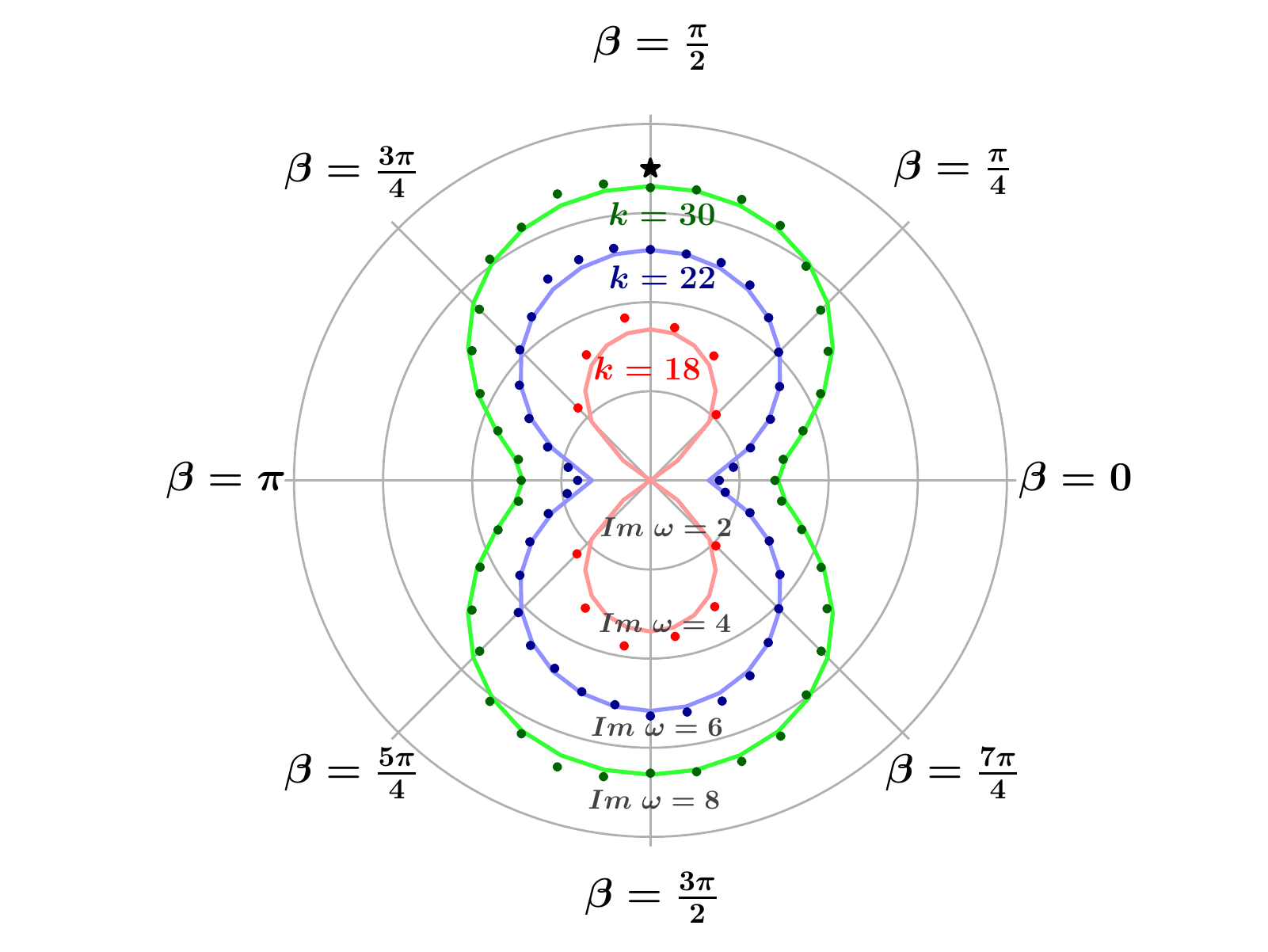}~
	\hspace{0.225 cm}
	\includegraphics[width=0.35\columnwidth]{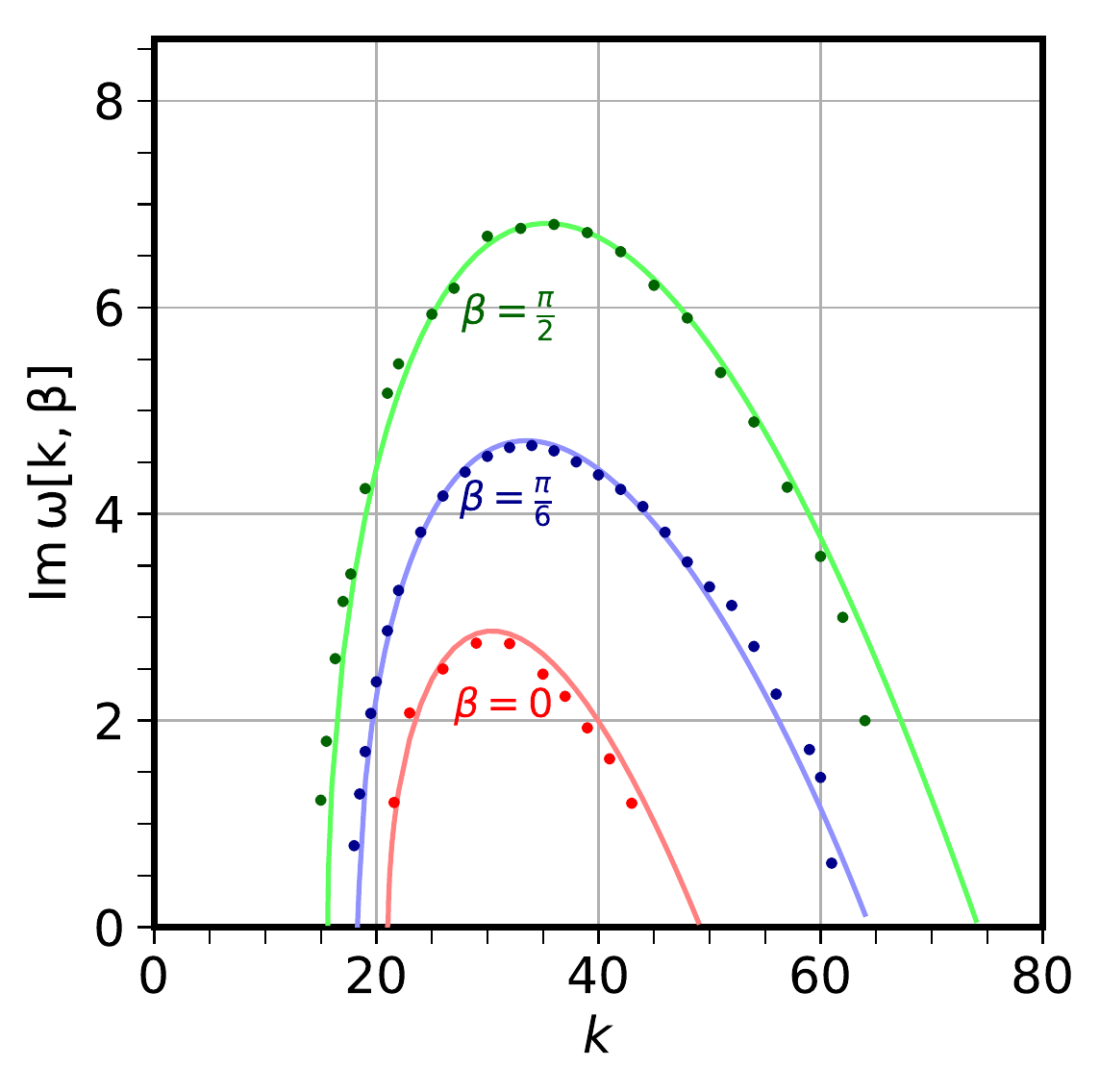}\\[1em]
	\textbf{Type\,II, A $\neq$ 0}\par\medskip
	\includegraphics[width=0.5\columnwidth]{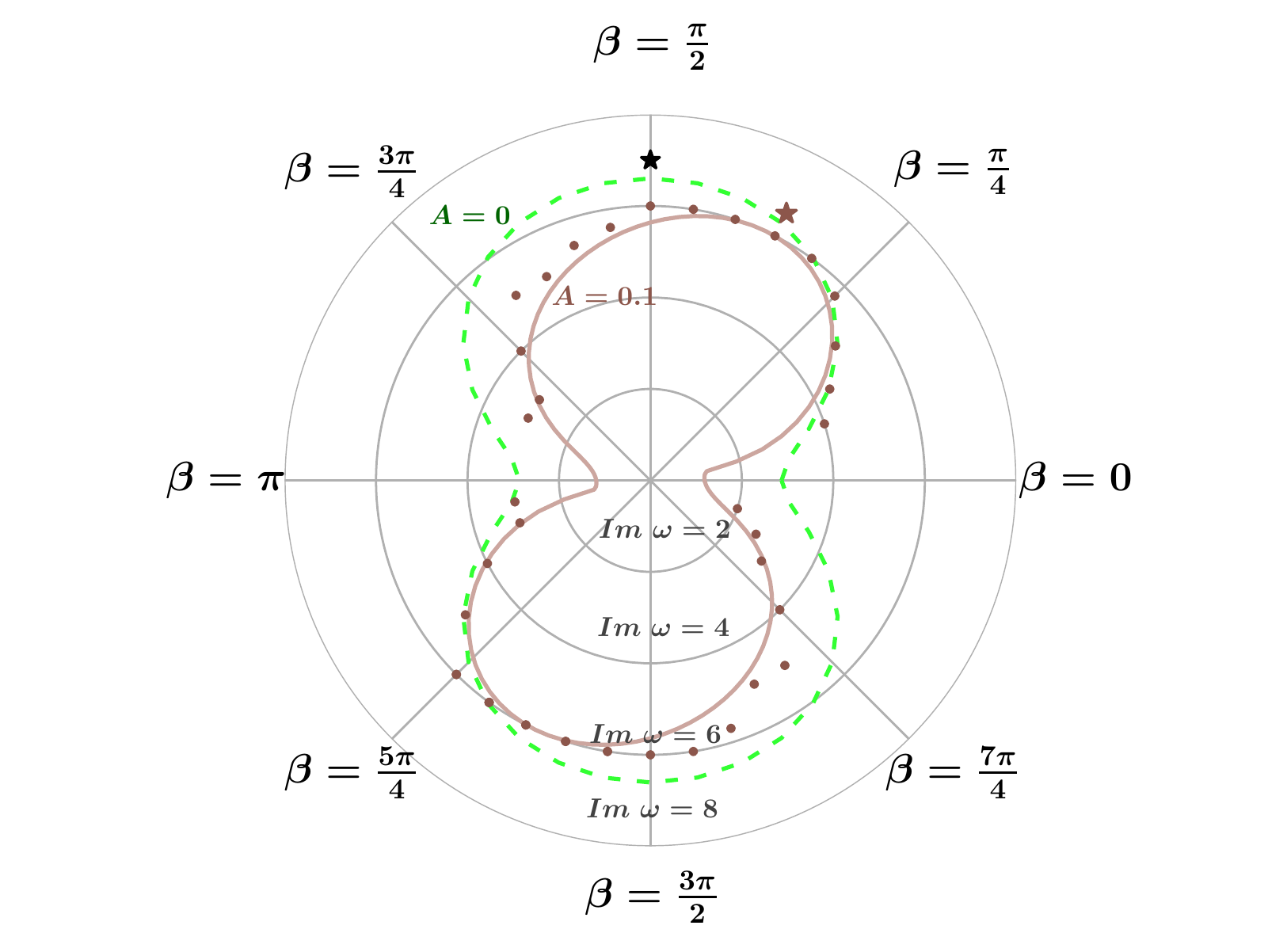}~
	\hspace{0.225 cm}
	\includegraphics[width=0.35\columnwidth]{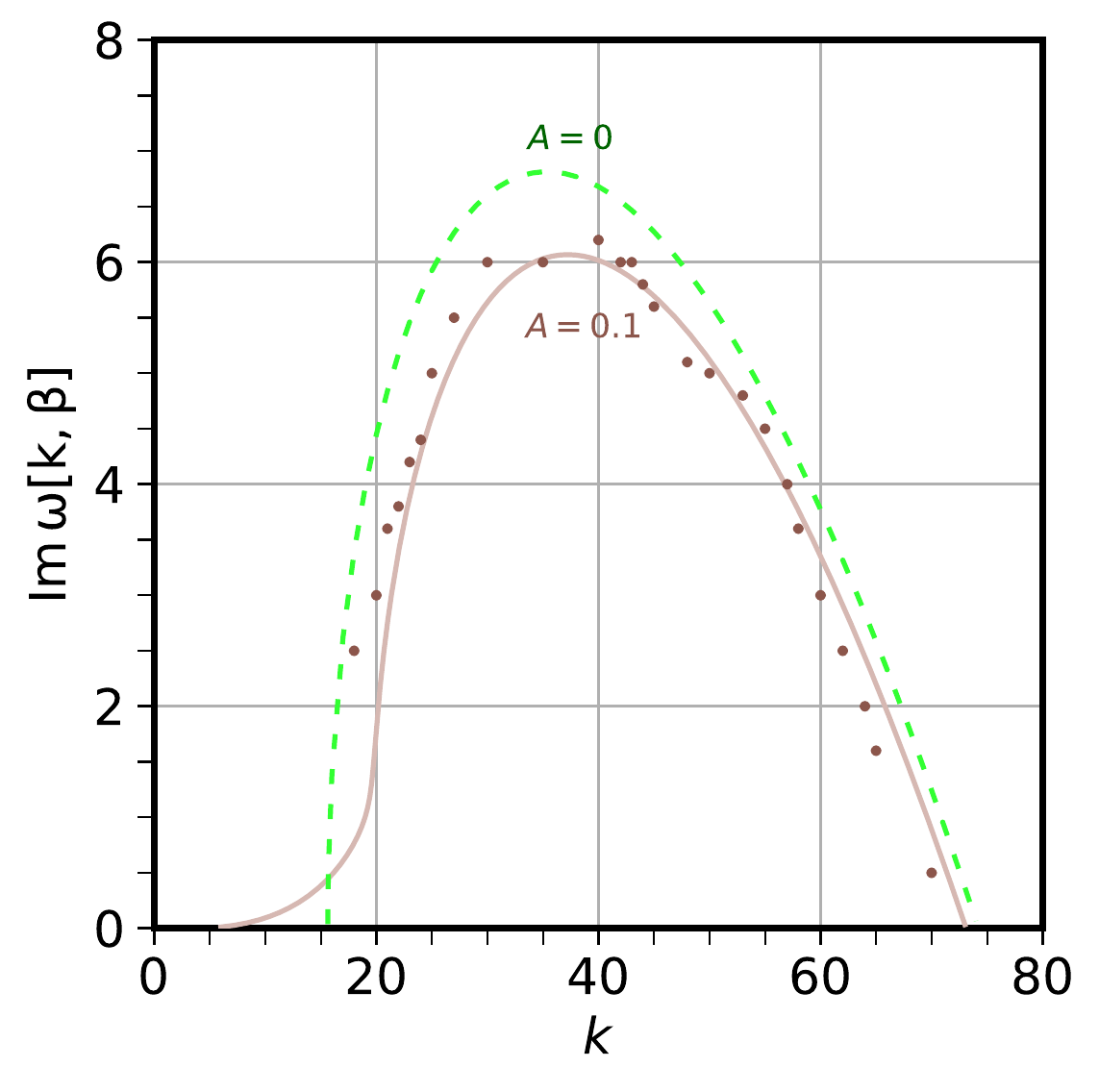}
\end{center}	
\caption{Left: Angular variation of $\rm{Im} \, \omega$ with respect to $\beta$. Right : Radial variation of $\rm{Im} \, \omega$ with respect to $k$. The top plots are done with $A = 0$ for different values of $k$ (left) and $\beta$ (right) whereas the bottom ones with fixed $k = 30$ (left) and $\beta = \frac{\pi}{2}$ (right) for $A = 0.1$. In these plots the continuous lines represent the linear stability solution while the dotted points the  solution of the full equation of motion. For comparison purpose the solution for $A = 0$ is also shown in green dashed lines in the bottom panel plots.} 
	\label{fig4}
\end{figure}
Type\,II ELNs have a symmetry under the joint operations $v_x \rightarrow -v_x$ and $v_y \rightarrow -v_y$, for any value of $A$. This results in a $\mathcal{D}[-k_x, -k_y, \omega] = \mathcal{D}[k_x, k_y, \omega]$ type of symmetry in $k_x-k_y$ plane. This can be understood through
\begin{equation}\label{21}
\begin{split}
{\Pi}_{ij}[k_{x}, k_{y}, \omega] &\xrightarrow [\text{$k_x \rightarrow -k_x$}]{\text{$k_y \rightarrow -k_y$}} {\eta}_{ij} + \int_{-1}^{1}\int_{-1}^{1} dv_{x} dv_{y} \hspace{1mm}\frac{G[{v_{x}, v_{y}}]}{\omega + k_xv_x+k_yv_y}\,\delta[v-1]\,{W}_{ij}[v_x, v_y]\\ &\xrightarrow [\text{$v_x \rightarrow -v_x$}]{\text{$v_y \rightarrow -v_y$}} \pm   {\Pi}_{ij}[k_{x}, k_{y}, \omega]\,.
\end{split}
\end{equation}
In the last step of Eq.\eqref{21}, $G[{-v_x, -v_y}] = G[{v_x, v_y}]$ has been used. Eq.\eqref{21} says that ${\Pi}_{ij}$ remains invariant under the joint operations of $k_x \rightarrow -k_x$ and $k_y \rightarrow -k_y$ except a minus sign that only occurs for $\Pi_{tx}$ and $\Pi_{ty}$. But interestingly again they come in pairs in Eq.\eqref{11}, such as $\left(\Pi_{tx}\right)^2, \left(\Pi_{ty}\right)^2$ or $\Pi_{tx}\Pi_{ty}$, leaving the dispersion relation as well as as the solution for $\rm{Im} \, \omega$ invariant under the above operations. An interesting special case occurs for Type\,II ELNs with $A = 0$ where now the dispersion relation can have two more symmetries. For example, $\mathcal{D}[-k_x, k_y, \omega] = \mathcal{D}[k_x, k_y, \omega]$ and $\mathcal{D}[k_x, -k_y, \omega] = \mathcal{D}[k_x, k_y, \omega]$ along with the previous one. These extra symmetries for $A = 0$ case can be easily understood using similar arguments as above.

Eq.\eqref{21} for $k_x = k_y = 0$ gives
\begin{equation}\label{24}
\Pi_{tx}[0, 0, \omega_{0}] = 0
\end{equation}
and 
\begin{equation}\label{25}
\Pi_{ty}[0, 0, \omega_{0}] = 0\,.
\end{equation}
This further simplifies Eq.\eqref{11} into two separate equations:
\begin{equation}\label{26}
\Pi_{tt}[0, 0, \omega_{0}] = 0 
\end{equation}
and 
\begin{equation}\label{27}
\Pi_{xy}[0, 0, \omega_{0}]\Pi_{xy}[0, 0, \omega_{0}] - \Pi_{xx}[0, 0, \omega_{0}]\Pi_{yy}[0, 0, \omega_{0}]= 0  \,.
\end{equation}

Eq.\eqref{26} implies $\omega_{0} = -\phi_{tt}$ which is a real solution. Eq.\eqref{27} simplifies to give rise to a quadratic equation,
\begin{equation}\label{28}
\omega_{0}^{2}-\omega_{0} \Bigl(\phi_{xx} + \phi_{yy}\Bigr) + \Bigl(\phi_{xx}\phi_{yy}-\left(\phi_{xy}\right)^{2}\Bigr) = 0\,.
\end{equation}
The solutions of Eq.\eqref{28} are complex only if
\begin{equation}\label{29}
\bigl(\phi_{xx} - \phi_{yy}\bigr)^{2}+4\bigl(\phi_{xy}\bigr)^{2} < 0\,. 
\end{equation}
The condition in Eq.\eqref{29} can never be fulfilled, as the LHS is a  sum of two perfect squares, thus implying a stable zero-mode for Type\,II ELNs. 

Fig.\,\ref{fig4} shows the angular variation of $\rm{Im} \, \omega$ predicted by our previous analytical arguments for this case. The stability of $k = 0$ mode as shown in the radial variation  in top-right plot of Fig.\ref{fig4}, confirms our analytical claim. Already one finds that having two crossings leads to less instability in some sense. Interestingly, the radial variation of $\rm{Im} \,\omega$ for this case shows an approximately Lorentzian shape as a function of $k$, i.e., large wavelength or small $k$ modes are inert then the growth rate increases as we increase $k$ with a maximum around $k = 30-40$ and then starts to decrease as a result very small wavelength or very large $k$ modes again become inert. The Lorentzian is much wider closer to the $\beta = \pi/2$ direction than at $\beta = 0$. The Fourier modes close to $k_{y}$ ($\beta = \pi/2$) axis have much larger growth rate compared to modes close to  $k_{x}$ ($\beta = 0$) for this case. In contrast with the Type\,I case, the Fourier mode with the largest growth rate in this case lies along the $k_{y}$ axis ($\beta = \frac{\pi}{2}$) about which there is a $k_y \rightarrow -k_y$ symmetry.

For $A\neq0$, the angular variation of $\rm{Im} \, \omega$ shown in the bottom left panel of Fig.\ref{fig4} shows a skewed symmetry. The non-zero value of lepton asymmetry breaks the symmetry along $k_{x}$ and $k_{y}$ axes keeping the symmetry along the  diagonals intact and also tilts the overall angular distribution towards one of the diagonals. The non-zero value of $A$ also shifts the position of the Fourier mode with the largest growth rate from the $k_{y}$ axis to along one of the diagonals. This plot also indicates that the Fourier modes along $\beta = \pi/2$ have a much larger growth rate compared to modes along $\beta = 0$ direction but with an exception that in this case the overall growth of the system is suppressed compared to zero lepton asymmetry case. The $k = 0$ mode is stable in this case also, as shown in bottom right plot of Fig.\,\ref{fig4}. The same plot also indicates that $\rm{Im} \, \omega$ as a function of $k$ for specific $\beta$ direction shows a similar Lorentzian nature but it is slightly shifted towards $k = 0$ with lower width compared to $A = 0$ case. The diagonal symmetry, the shift in the position of the maximum of  $\rm{Im} \, \omega$ and the decrease in overall growth of the system become more pronounced as we increase the value of lepton asymmetry.

\subsection{Many Crossings}
\label{sec:multiple}
\begin{figure}
\begin{center}
	\includegraphics[width=0.45\columnwidth]{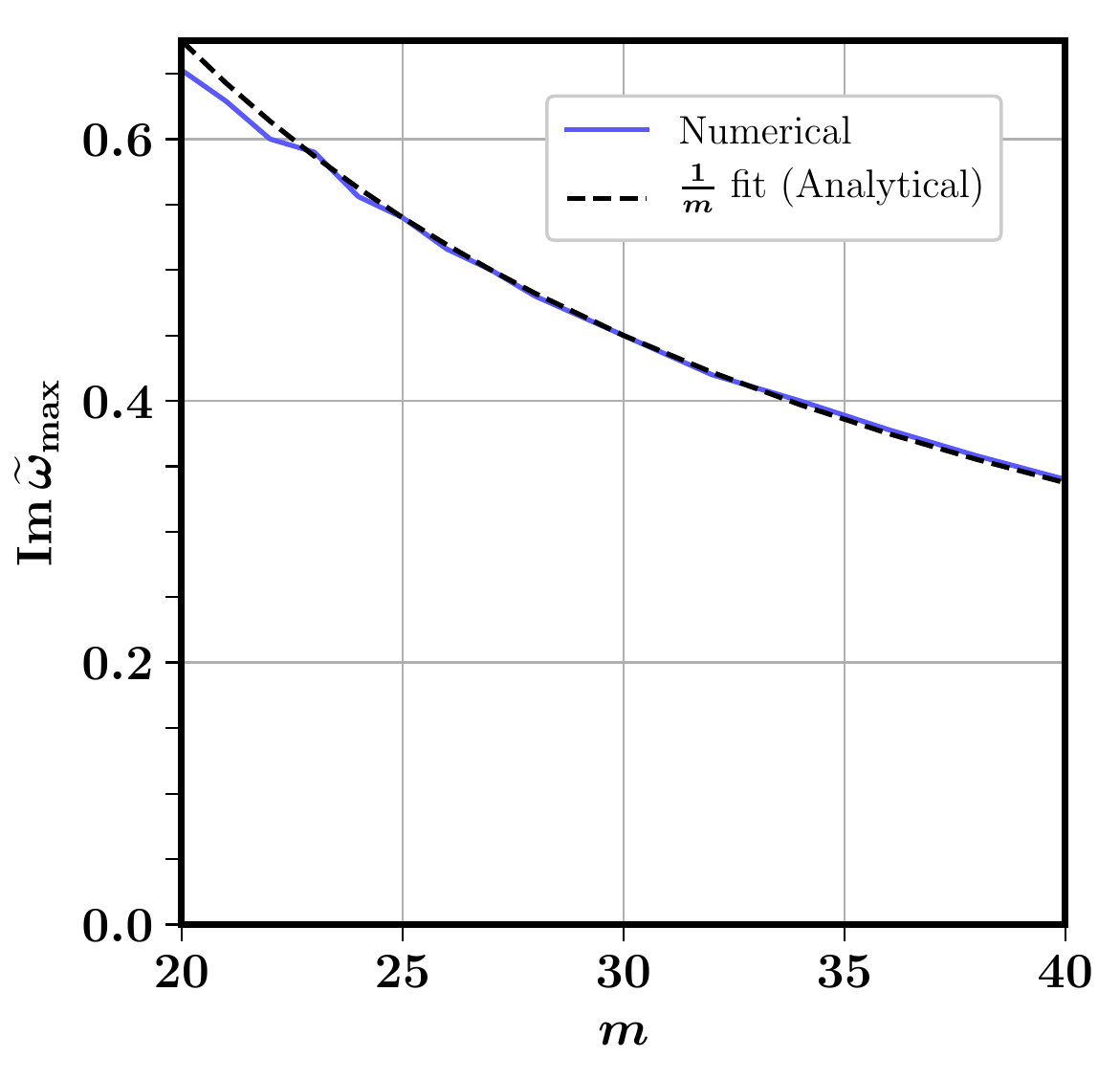}~
\end{center}	
\caption{${\rm Im}\, \widetilde{\omega}_{\rm max}$ as a function of the crossing number $m$ is shown for $G[\theta] = 6 \sin{m\theta}$. Analytical  arguments show that ${\rm Im}\, \widetilde{\omega}_{\rm max}$ decreases as $1/m$.} 
\label{fig5}
\end{figure}

Type\,III ELN corresponds to a scenario where, e.g., due to fluctuations of the neutrino distributions, the angular distributions of $\nu_{e}, \overline{\nu}_{e}$ are rapidly changing as a function of $\theta$ and their difference can go through large number of zero crossings. To mimic this scenario we considered the Type\,III $G[\theta]$ with $m$ being quite large. We will show how the maximum growth of such systems depends on the number of zero crossings, closely related to $m$. To understand this let us consider Eq.\eqref{6} in terms of the angular variable $\theta$ and the $k-\beta$ coordinates, 
\begin{equation}\label{30}
\mathbf{\Pi}[k, \beta, \omega] = \boldsymbol{\eta} + \boldsymbol{\psi}[\omega, k, \beta]= \boldsymbol{\eta}+ \int_0^{2\pi} d\theta \hspace{1mm} f[\theta, \omega, k, \beta ]\,G[\theta]\,\mathbf{W}[\theta]\,,
\end{equation}
where we have defined $\boldsymbol{\psi}[\omega, k, \beta]$ as the matrix of integrals on the RHS. As before, but now explicitly in terms of $\theta$, one has
\begin{equation}\label{33}
f[\theta, \omega, k, \beta] = \frac{1}{\omega-k \cos{\left(\theta -\beta\right)}}
\end{equation}
and 
\begin{equation}\label{32}
\mathbf{W}[\theta]  = \begin{pmatrix}
1 & \cos{\theta} & \sin{\theta} \\
\cos{\theta} & \cos^2{\theta} & {\sin{\theta}}\cos\theta \\
\sin{\theta}  & {\sin{\theta}}\cos\theta   & \sin^2{\theta} 
\end{pmatrix}\,.
\end{equation}
Eq.\eqref{33} indicates that the integrands in each component of $\boldsymbol{\psi}[\omega, k, \beta]$ has 
a saddle-point at say $\theta_0[i, j]$. In general, depending on ${W}_{ij}[\theta]$, the stationary points $\theta_0[i,j]$ will be shifted away from $\cos^{-1} {\left(\frac{{\rm Re}\,\omega}{k}\right)} + \beta$, and will not be the same for the different $\left(i, j\right)$ components. But from now on for convenience we will stop explicitly writing the functional dependence on $i, j$ in $\theta_0$. Our objective, will  be to compute the integral in Eq.\eqref{30} in the saddle-point approximation, and show how the Fourier mode with maximum growth, labeled by $\left(\omega_{\rm max}, k_{\rm max}, \beta_{\rm max}\right)$, has smaller instability for larger $m$.

First we rewrite the $ij^{\rm th}$ component of $\boldsymbol{\psi}[\omega_{\rm max}, k_{\rm max}, \beta_{\rm max}]$ in terms of the logarithm of its integrand as,
\begin{equation}\label{34}
\psi_{ij}[\omega_{\rm max}, k_{\rm max}, \beta_{\rm max}]=\int_0^{2\pi} d\theta\,f[\theta, \omega_{\rm max}, k_{\rm max}, \beta_{\rm max}]\,G[\theta]{W}_{ij}[\theta] =\int_0^{2\pi} d\theta\, \exp[{F_{ij}[\theta]}]\,. 
\end{equation}
We then expand $F_{ij}[\theta]$ about its stationary point $\theta_0$ up to second order to obtain,
\begin{equation}\label{35}
F_{ij}[\theta] = F_{ij}[\theta_0]+\left(\theta-\theta_0 \right)^2 \frac{d^2 F_{ij}[\theta]}{d\theta^2} \bigg | _{\theta=\theta_0}\,.
\end{equation}
Note $F_{ij}[\theta]$ here is a complex function. Eq.\eqref{34} and Eq.\eqref{35} allow us to perform the saddle-point integral around $\theta_0$ to get
\begin{equation}\label{36}
\psi_{ij}[\omega_{\rm max}, k_{\rm max}, \beta_{\rm max}] = f[\theta_0, \omega_{\rm max}, k_{\rm max}, \beta_{\rm max}]\,G[\theta_0]\,{W}_{ij}[\theta_0]\,\sqrt{\frac{2\pi}{-\frac{d^2 F_{ij}[\theta]}{d\theta^2}\big| _{\theta=\theta_0}}}\,.
\end{equation}
Now, one can insert the expression of Type III ELN and use the large-$m$ limit to write
\begin{equation}\label{37}
-\frac{d^2 F_{ij}[\theta]}{d\theta^2}\bigg| _{\theta=\theta_{0}} = \frac{m^2\left(c_1^2+c_2^2+\frac{A}{2\pi} G[\theta_0]-\frac{A^2}{4\pi^2}\right)}{G[\theta_0]^2}+{\cal O}(m^0)\,,
\end{equation}
where $c_{1,2}$ are the coefficients of the $\sin$ and $\cos$ terms in the Type\,III ELN.
The ${\cal O}(m^0)$ can be neglected for Type\,III ELNs using the fact that $m$ is large. Eq.\eqref{37} allows us to approximately simplify Eq.\eqref{34} to
\begin{equation}\label{38}
{\psi}_{ij}[\omega_{\rm max}, k_{\rm max}, \beta_{\rm max}] = \widetilde{G}[\theta_0] \frac{{W}_{ij}[\theta_0]}{m\, \widetilde{\omega}_{\rm max}}\,,
\end{equation}
where 
\begin{equation}
\widetilde{G}[\theta_0] = \frac{{\sqrt{2\pi}}}{{\left({c_1^2+c_2^2+\frac{A}{2\pi} G[\theta_0]-\frac{A^2}{4\pi^2}}\right)^{1/2}}}\,G^2[\theta_0]\,,
\end{equation}
and 
\begin{equation}
{\widetilde{\omega}_{\rm max}}={\omega_{\rm max} - k_{\rm max} \cos{\left(\theta_0-\beta_{\rm max}\right)}}
\end{equation}
is the complex growth rate for the mode $({k}_{\rm max},\beta_{\rm max})$.  Note ${\widetilde{\omega}_{\rm max}}$ in principle can have dependence on $i, j$ indices via $\theta_0$ but as it appears only in the real part of ${\widetilde{\omega}_{\rm max}}$ and in the limit $\theta_0 [i, j]$ for different $i, j$ indices are close to $\cos^{-1} {\left(\frac{{\rm Re}\,\omega_{\textrm{max}}}{k_{\textrm{max}}}\right)} + \beta_{\textrm{max}}$ or $\textrm{Re} \, {\widetilde{\omega}_{\rm max}} \approx 0$, it can be ignored.

Equation \eqref{38} shows that $\widetilde\omega_{\rm max}$, and thus $\omega_{\rm max}$, always appears multiplied by $m$. It is then obvious that ${\rm Im}\,\omega_{\rm max}$ must scale as $1/m$. We checked this behavior by numerically solving the dispersion relation and then locating the maximum of this solution in the $k-\beta$ plane. For Type III ELN with $A = 0, c_1 = 1, c_2 = 6$, and $m$ in the range $20-40$ for which the ELN has many zero crossings, this numerical result is shown as the blue continuous line in Fig.\,\ref{fig5}. The black dashed line is the best fit of that numerical data with $\frac{a_0}{m}+a_1$ where the fitted parameter values are $\left(a_0, a_1 \right) = \left(13.5, 0.0\right)$. This behavior clearly supports our analytical claim that a large number of crossings leads to the instability growth being hindered as $1/m$. This is of course obtained with a very particular form of the Type III ELN, which is purely sinusoidal with a single $m$.  Based on numerical experiments, we conjecture that growth rates with many crossings should be small in general, all else being equal.
Generalizing the above analysis for an arbitrary ELN, say written using a sine and cosine series in $\theta$, does not seem straight-forward.

\section{Summary}
\label{sec:conc}
In this paper we explored, analytically and numerically, how the initial growth of flavor instabilities of a dense neutrino gas depends on the number of zero crossings in the ELN. Improving upon previous lower-dimensional studies, this is the first study in 2 (space) + 2 (momentum) + 1 (time) dimensions. We developed our own code that solves for the flavor evolution of dense neutrinos. We also presented a new strategy to solve the transcendental equations appearing in the dispersion relation for the linear evolution of such systems. With these new tools we explored the linear behavior by looking at the different radial and angular distributions of the dispersion relation in the $k-\beta$ plane. Our main results are
\begin{itemize}
\item The symmetries of the ${\rm Im}\,\omega$ in the $k-\beta$ plane, its radial (i.e., vs. $k$) and angular (i.e., vs $\beta$) variation, the stability of the $k = 0$ mode, overall linear growth and the position of the Fourier mode with the highest growth rate, etc., all have an intimate connection with the various symmetries of the neutrino angular distributions and one can analytically understand them in great detail. Figs.\,\ref{fig3}, \ref{fig4}, \ref{fig5} show the exquisite match between growth rates predicted by linear stability analysis and fully numerical evaluation of the solutions. This matching and understanding over a variety of lepton asymmetries, $A$, shapes of ELNs, different wavelength $k$ and directions $\beta$, shows the extraordinary power of linear theory and a testament to the fidelity of our numerics.

\item ELNs with large number of zero crossings lead to a relatively smaller growth rate, essentially decreasing as $1/m$ where $m$ is the number of crossings. We speculate that this may be important for many realistic environments where $\nu_e$ and $\bar\nu_e$ distributions are close to each other and  crossings occur in the ELNs due to noise or fluctuations. It seems that the growth rates for such instabilities will be relatively suppressed.
\end{itemize}

\section*{Acknowledgements}
The work of B.D. is supported by the Dept.\,\,of Atomic Energy (Govt.\,\,of India) research project under Project Identification No. RTI 4002, the Dept.\,\,of Science and Technology (Govt.\,\,of India) through a Swarnajayanti Fellowship, and by the Max-Planck-Gesellschaft through a Max Planck Partner Group.

\appendix
\section{Error Estimate for Numerical Solutions}
\label{sec:appendix}
\begin{figure*}[]
\centering
	\includegraphics[width=0.4\columnwidth]{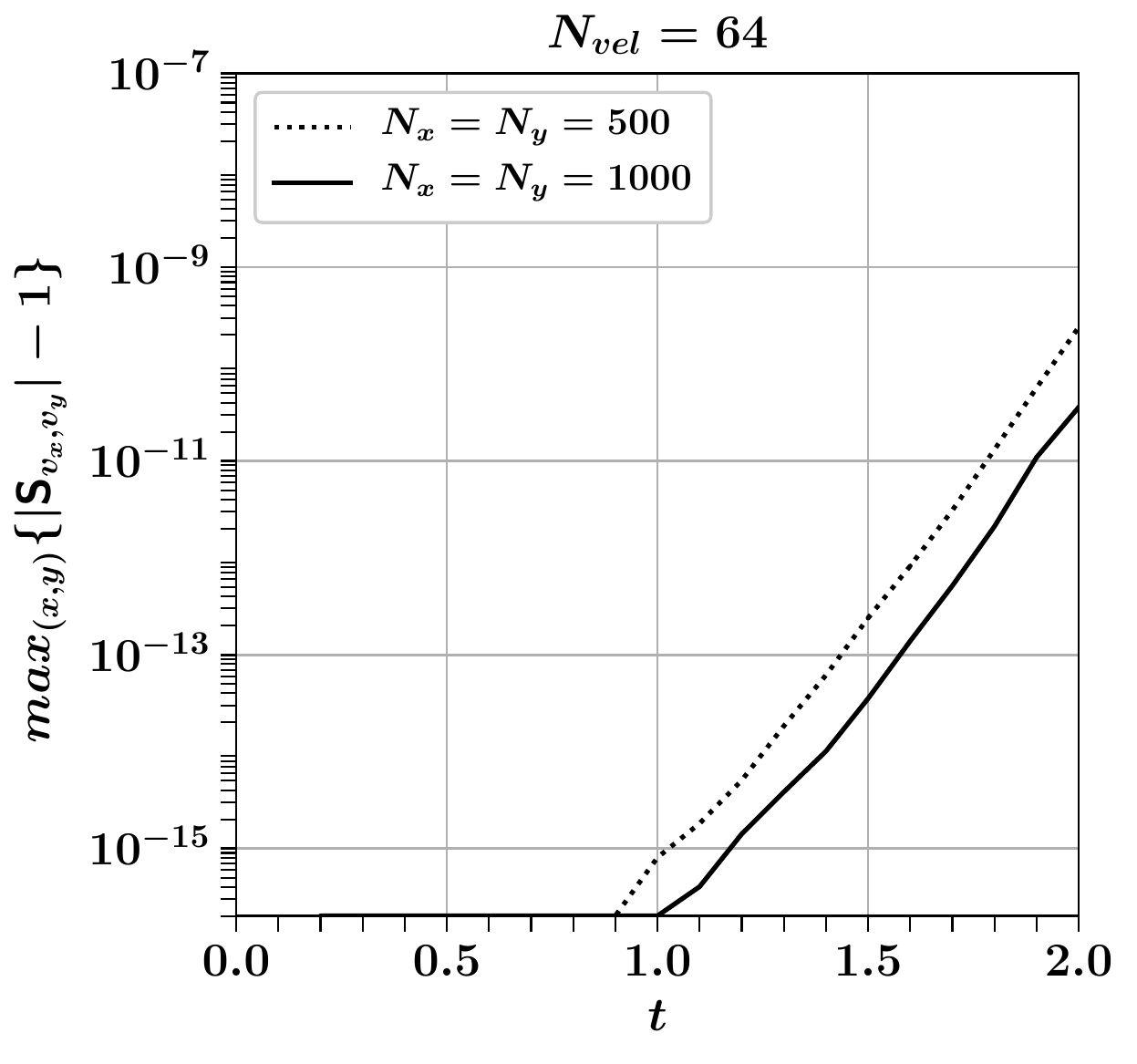}~\quad~\quad
	\includegraphics[width=0.4\columnwidth]{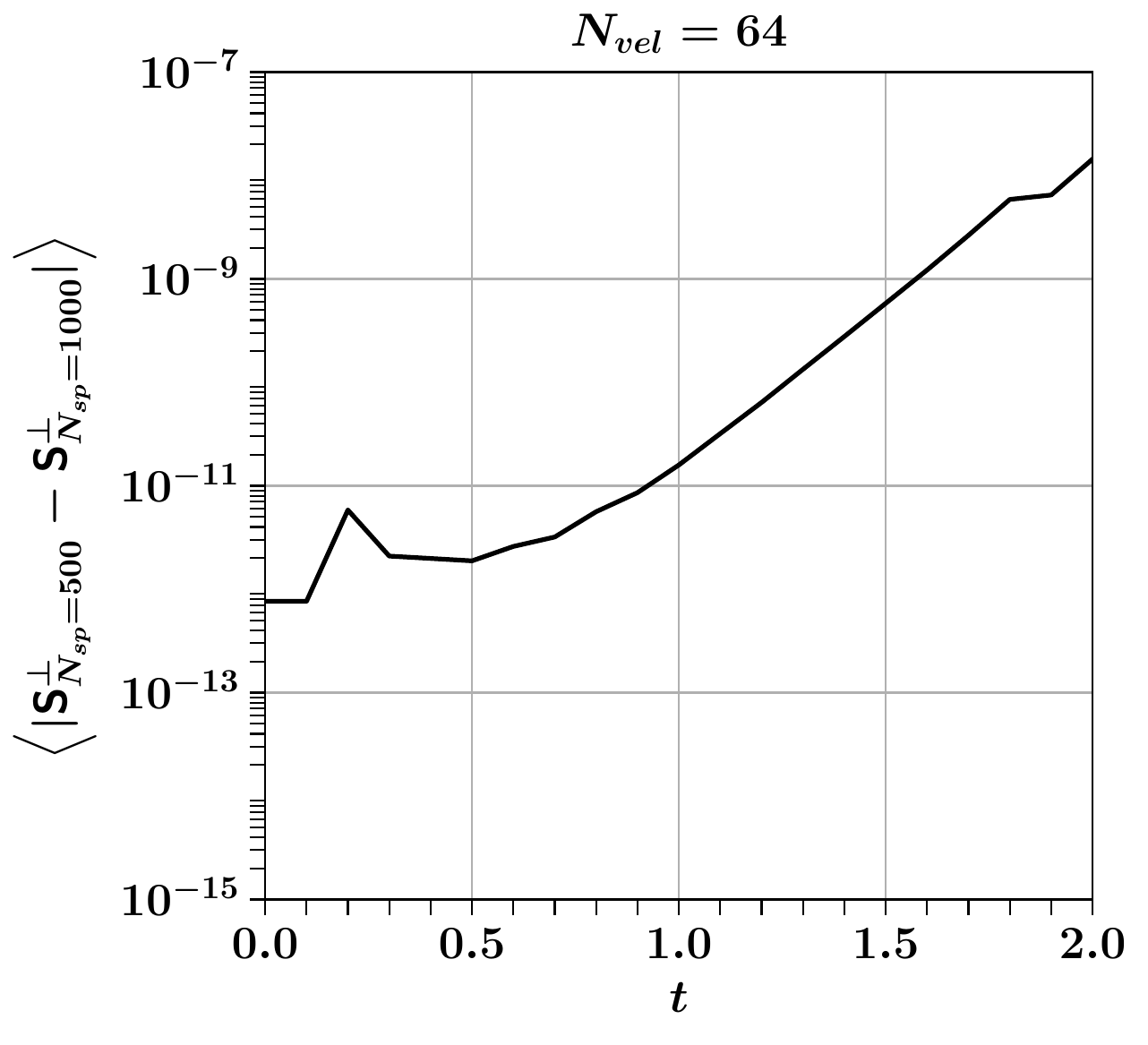}~
	\caption{Left: Accuracy of our calculation, estimated by maximum departure of $|{\mathsf{S}}|$ from unity in the $x-y$ plane, as a function of time $t$. Right: Precision of our calculation, estimated by convergence of $\langle{|\mathsf{S}}^\perp|\rangle$ with respect to our best discretization, i.e., $\left(N_x = N_y = 1000\right)$, shown as a function of time $t$. In all these calculations, we choose $N_{vel} = 64$ and for these plots we show the mode $\{v_{x} = 0.87, v_{y} = 0.5\}$. } 
	\label{figA}
\end{figure*}

\begin{figure*}[]
	\includegraphics[width=0.3\columnwidth]{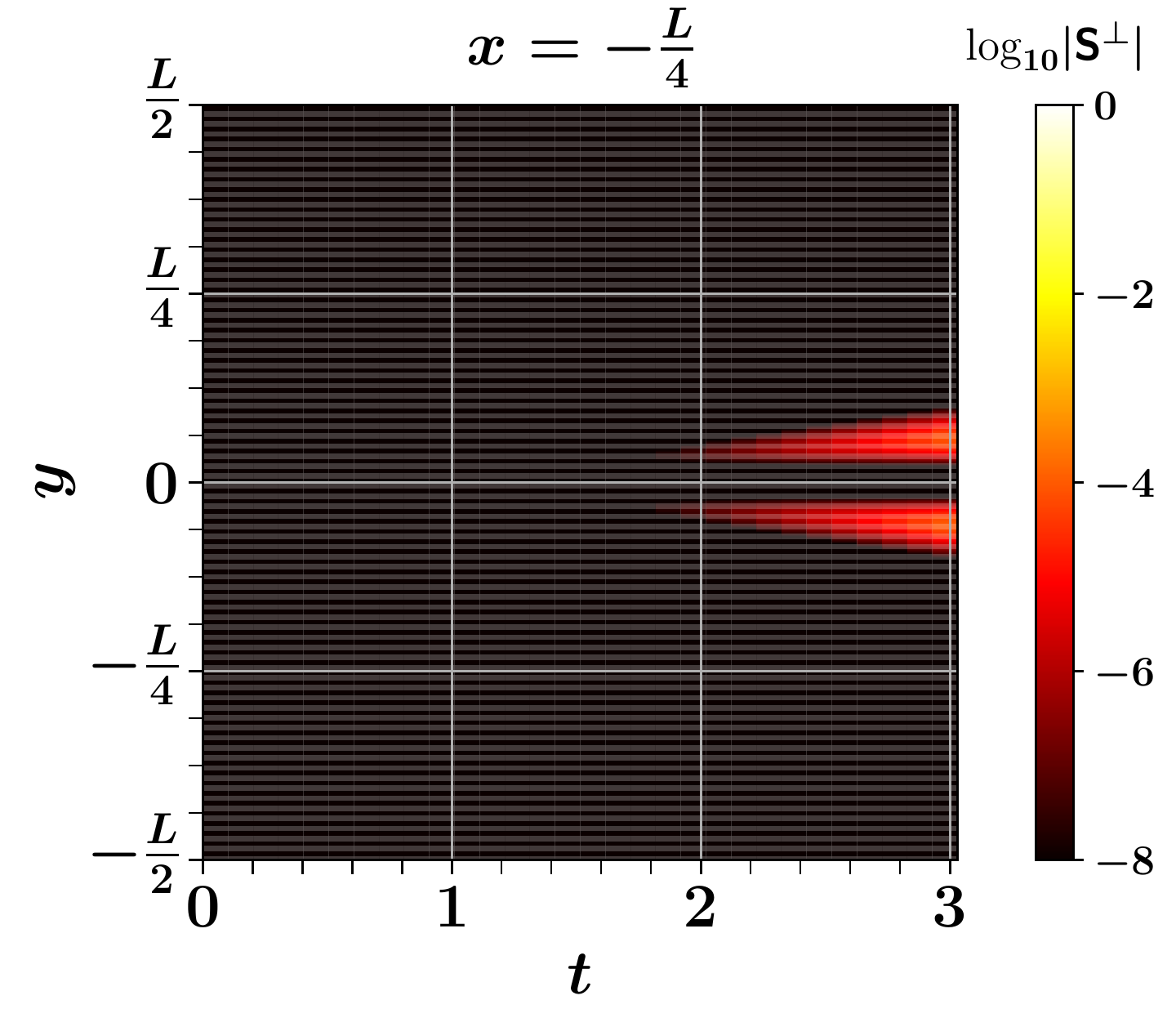}~
	\includegraphics[width=0.3\columnwidth]{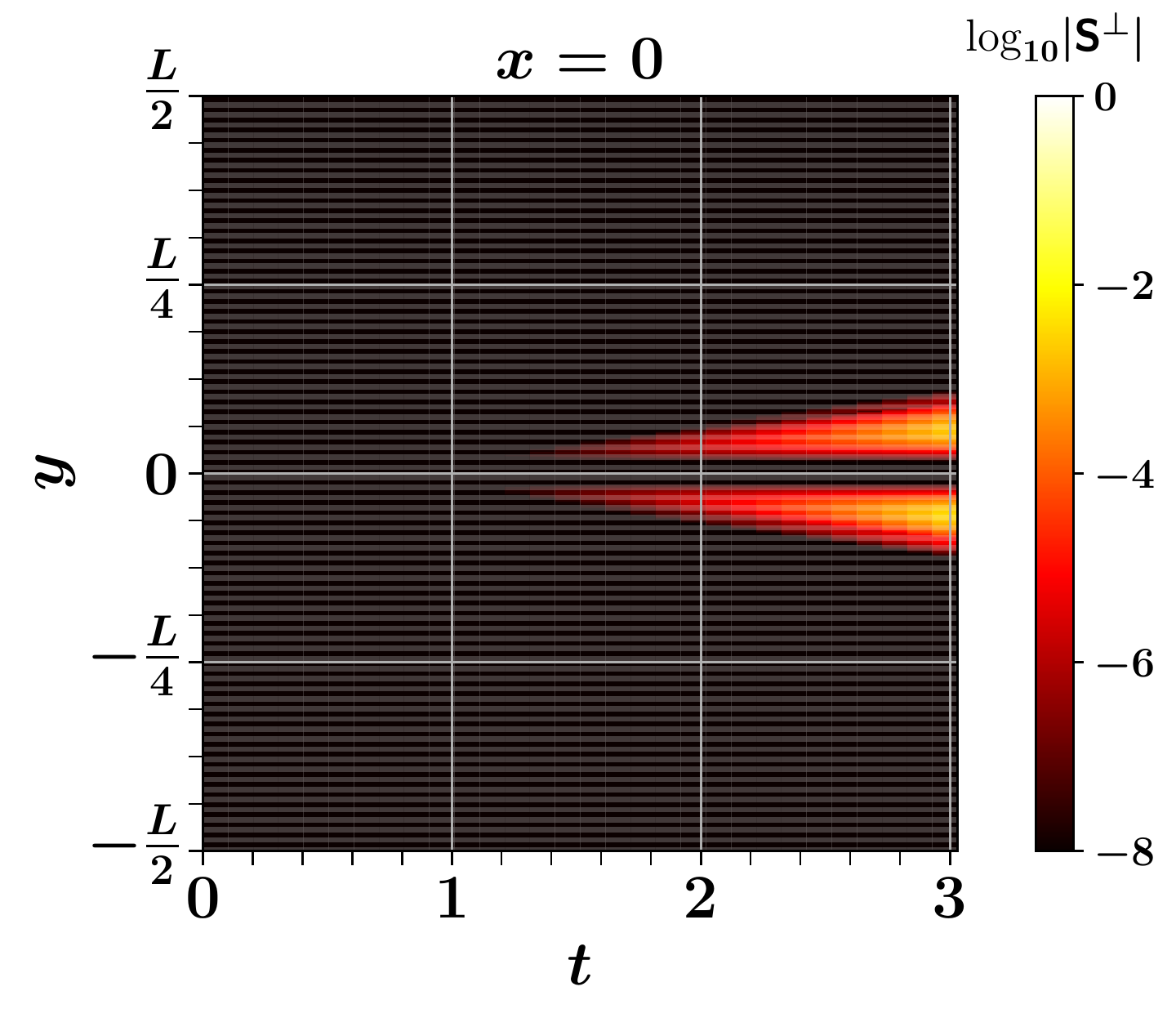}~
	\includegraphics[width=0.3\columnwidth]{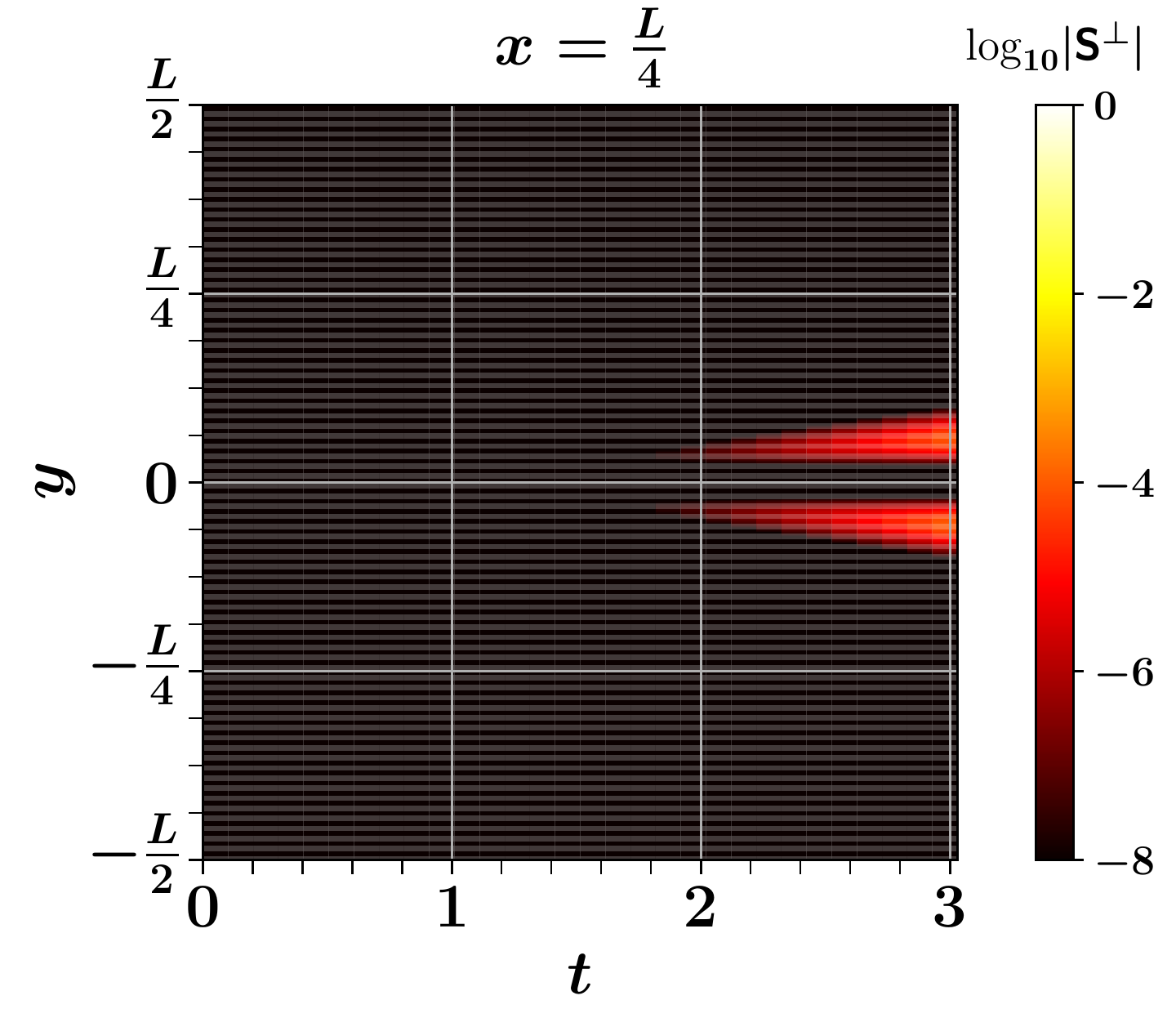} \\
	\includegraphics[width=0.3\columnwidth]{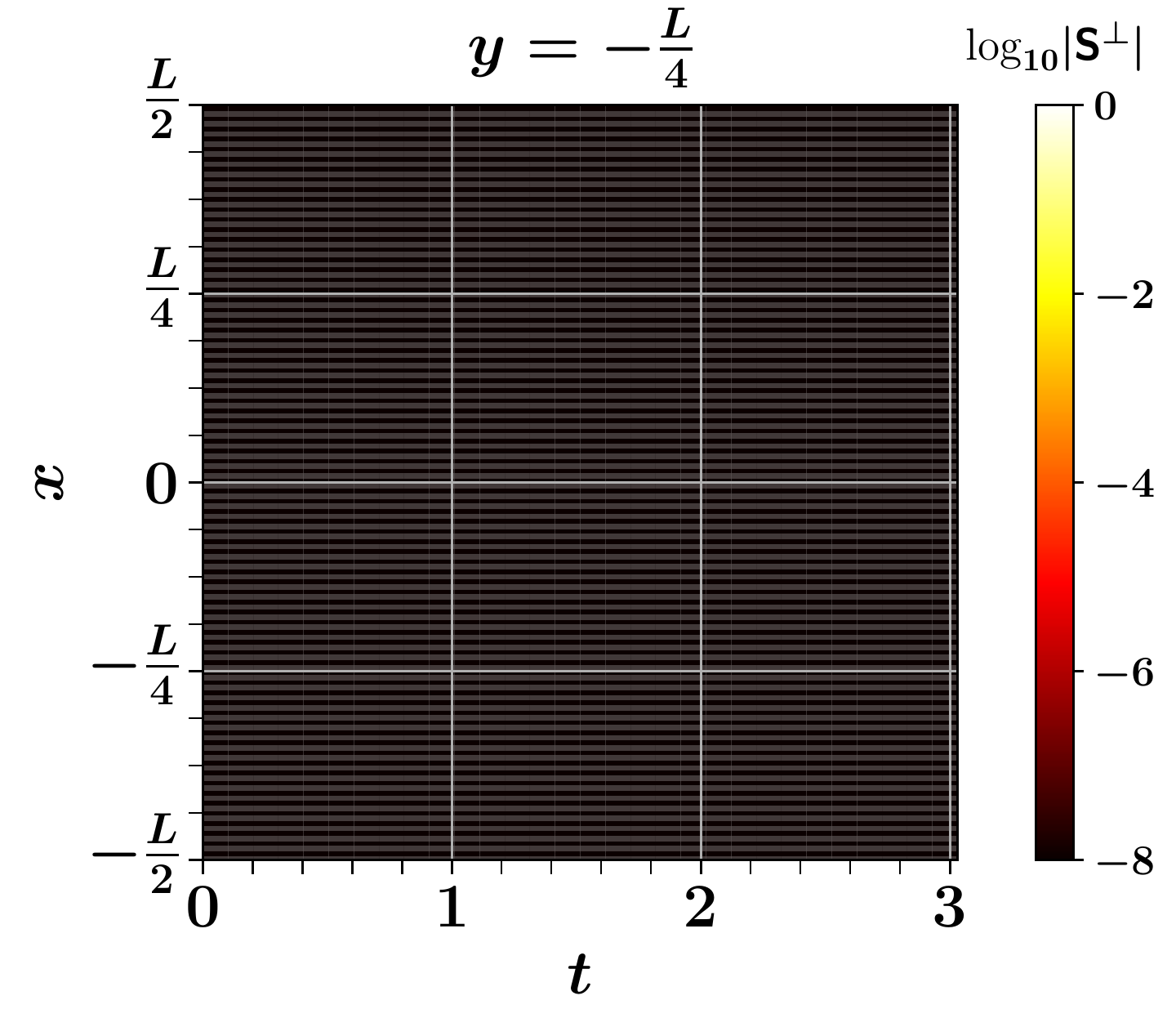}~
	\includegraphics[width=0.3\columnwidth]{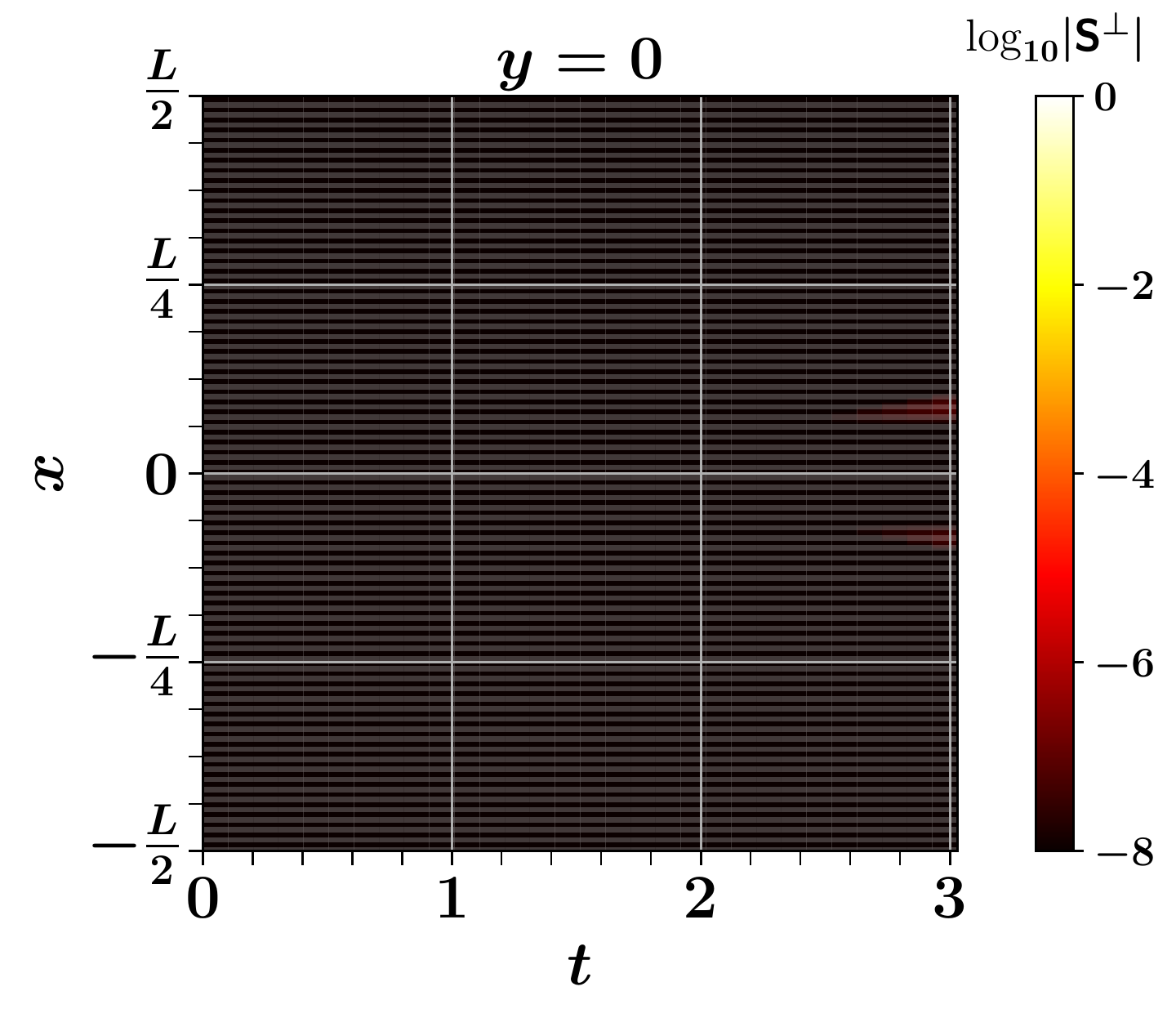}~
	\includegraphics[width=0.3\columnwidth]{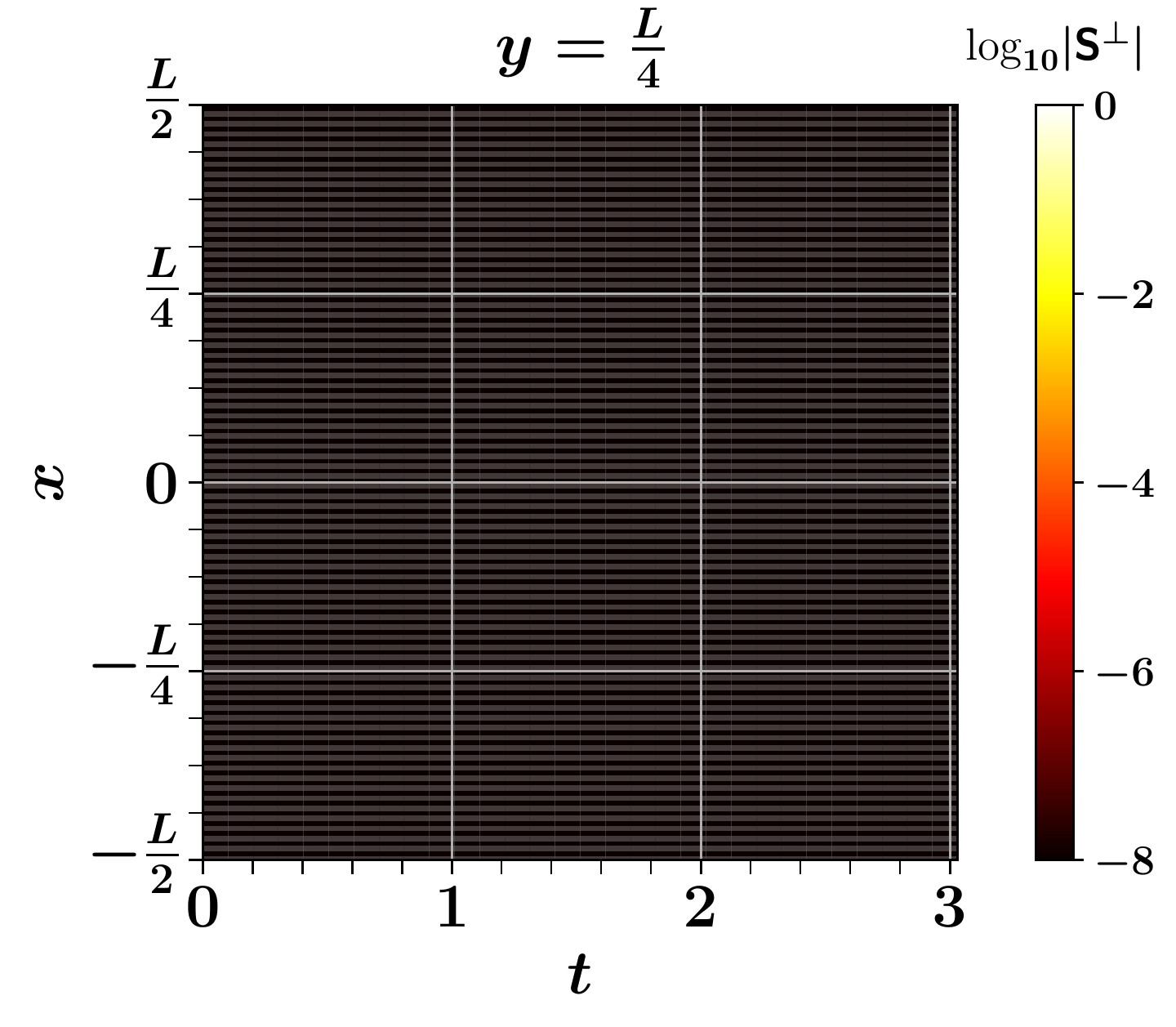} 
	\caption{Top : Flavor evolution in the $y-t$ plane obtained from the numerical solution of Eq.\eqref{2} for Type\,II ELNs with $A = 0.1$, shown at three different $x$ positions. 
		Bottom : Same in the $x-t$ plane at three $y$ positions. The color-coding in the colorbar depicts the log$_{10}$ of the magnitude of $\mathsf{S}^{\perp}$.} 
	\label{figB}
\end{figure*}
In Fig.\,\ref{figA} (left panel), we show the  accuracy expected of our calculation. We check if the length of the polarization vectors remain fixed at unity or not. The error we incur on this is a lower bound on the error in our calculations. We find that our chosen discretization, $N_x=N_y=500$ and $N_v=64$, does as well as finer discretizations, incurring an error of ${\cal O}(10^{-10})$ at $t \approx 2$ where the linear growth of the system ends. Even in the far nonlinear regime, the error remains well under $10^{-3}$. For illustration, here we have chosen a Type\,II ELNs with nonzero lepton asymmetry.

In Fig.\,\ref{figA} (right panel), we illustrate the precision to be expected of our numerical solutions of the equations of motion. We check for convergence by comparing the length of the spatially averaged version of the polarization vector perpendicular to the $z$-axis between two different discretizations: one with $N_x = N_y = 500$ and the other $N_x = N_y = 1000$. The computations are shown with $N_{vel} = 64$. Our results indicate that a discretization of $N_x= N_y= 500$ is at most ${\cal O}(10^{-8})$ off from yet finer discretizations in the linear regime which ends almost at $t = 2$. This result is also shown for the same velocity mode with the same choice of ELN.

For completeness, we show the flavor evolution on the $x-t$ plane (resp. $y-t$ plane) at three $y$ (resp. $x$) positions for the above-considered case in Fig.\,\ref{figB}. Note the overall growth in flavor along $y$ direction is much larger compared to $x$ direction and it decreases from the center towards the edge of the box in either directions, as dictated by the fastest growing $\vec{k}$ mode. The box has been chosen to be much larger than the region where the solution is nontrivial; this is to avoid artifacts of the finiteness of the box.

\bibliographystyle{JHEP}
\bibliography{references}

\end{document}